\documentclass[12pt,a4paper]{article}
\usepackage[utf8]{inputenc}
\usepackage[english]{babel}
\usepackage{amsmath}
\usepackage{amsfonts}
\usepackage{amssymb}
\usepackage{amsthm}
\usepackage{multirow}
\usepackage{graphicx}
\usepackage{bbm}
\usepackage{enumerate}
\usepackage[left=2cm,right=2cm,top=2cm,bottom=2cm]{geometry}
\usepackage{natbib}
\usepackage{color}
\usepackage{url}

\newtheorem{thm}{Theorem}

\newcommand{\refmath}[1]{(\ref{#1})}
\newcommand{\refmathanc}[1]{(M\ref{#1})}	% command used to cite the ANCOVA assumptions
\newcommand{\refmathmancats}[1]{(M\ref{#1})$^{\ast}$}	% command used to cite the MANCATS assumption
\newcommand{\refmathm}[1]{(M\ref{#1})}	% command used to cite the MANCATS assumptions

\newcommand{\mf}[1]{{\bf #1}}
\newcommand{\mfg}[1]{{\bf #1}}
\newcommand{\vf}[1]{{\bf #1}}
\newcommand{\vfg}[1]{\boldsymbol{#1}}
\newcommand{\best}{\vfg{\hat{\beta}}}

\newcommand{\bpar}{\vfg{\beta}}
\newcommand{\SigmaFull}{\mfg{\Sigma}_N}			% this is the true covariance matrix
\newcommand{\SigmaWhiteFull}{\mfg{\hat{\Sigma}}_N}	% This is the White HCCME
\newcommand{\xm}{\mf{X}}
\newcommand{\ip}{\mf{I}_p}
\newcommand{\dconv}{\overset{d}{\longrightarrow}}
\newcommand{\pconv}{\overset{P}{\longrightarrow}}

\allowdisplaybreaks[2]

\begin{document}

\title{Multivariate analysis of covariance when standard assumptions are violated}

\author{Georg Zimmermann\thanks{Georg Zimmermann, Department of Mathematics, Paris Lodron University, Hellbrunner Strasse 34, A 5020 Salzburg, Austria (georg.zimmermann@pmu.ac.at)}\\
Department of Mathematics, Paris Lodron University, Salzburg, Austria \\
Department of Neurology, Paracelsus Medical University, Salzburg, Austria\\
Spinal Cord Injury and Tissue Regeneration Centre Salzburg,\\ Paracelsus Medical University, Salzburg, Austria\\
	\and Markus Pauly \\ Department of Statistics, Ulm University, Germany \\
	Institute for Mathematical Statistics and Applications in Industry, \\TU Dortmund, Dortmund, Germany\\
	\and Arne C. Bathke \\
	Department of Mathematics, Paris Lodron University Salzburg, Austria \\
	Department of Statistics, University of Kentucky, Lexington, KY, USA}

\date{}

\maketitle

\begin{abstract}
In applied research, it is often sensible to account for one or several covariates when testing for differences between multivariate means of several groups. However, the ``classical'' parametric multivariate analysis of covariance (MANCOVA) tests (\textit{e.g.,} Wilks' Lambda) are based on quite restrictive assumptions (homoscedasticity and normality of the errors), which might be difficult to justify in small sample size settings. Furthermore, existing potential remedies (\textit{e.g.,} heteroskedasticity-robust approaches) become
%might be 
inappropriate in cases where the covariance matrices are singular. Nevertheless, such scenarios are frequently encountered in the life sciences 
%medical research, psychology, 
and other fields, when for example, in the context of standardized assessments, a summary performance measure as well as its corresponding subscales are analyzed. 
%Therefore, 
In the present manuscript, we 
%extend some ideas which have been proposed for multivariate analysis of variance models and 
consider a general MANCOVA model, allowing for potentially heteroskedastic and even singular covariance matrices as well as non-normal errors. We combine heteroskedasticity-consistent covariance matrix estimation methods with our proposed modified MANCOVA ANOVA-type statistic (MANCATS) and apply two different bootstrap approaches.  We provide the proofs of the asymptotic validity of the respective testing procedures as well as the results from an extensive simulation study, which indicate that especially the parametric bootstrap version of the MANCATS outperforms its competitors in most scenarios, both in terms of type I error rates and power. These considerations are further illustrated and substantiated by examining real-life data from standardized achievement tests.  
\end{abstract}


\section{Introduction}
\label{Section:Intro}
%{\color{red} Bzgl. Journal: JMVA w\"are nat\"urlich quasi der Klassiker, und vielleicht hat es der Editor ja gerne, wenn er eine gewisse Weiterentwicklung innerhalb des Journals sieht (MATS $\rightarrow$ MANCATS). Aber nat\"urlich sind die anderen beiden Journale auch sehr nett – was meint ihr? \\ Arne: JMVA waere einen Versuch wert...}
Consider a randomized clinical trial, where each subject is randomized to one out of $a$ groups. Frequently, multiple outcomes (\textit{e.g.,} one primary and one co-primary outcome, several secondary outcomes) need to be compared between the groups. One straightforward way of accomplishing this goal would be to conduct univariate tests (\textit{e.g.,} $t$ tests) and adjust for multiplicity. Nevertheless, this approach has the drawback of not taking potential correlations between the outcomes (\textit{e.g.,} due to physiological reasons, there might be correlations between laboratory measurements) into account. Moreover, when analyzing global assessment variables, further safety and efficacy outcomes, or objective variables related to the global assessment, should be analyzed \citep{IchE9}. In these instances, employing a multivariate analysis of variance model would be an attractive option. Still, however, the estimators of the treatment effects might be biased, and the power of the test might be low, if covariates (\textit{i.e.,} variables that are thought to influence the outcomes) are not included in the model. Therefore, a multivariate analysis of covariance (MANCOVA) model might be a better choice. Indeed, MANCOVA models have been used in several recent publications in medical research \citep{Rol13, Jac18, Set18, Tou18} and psychology \citep{Mem16,Lyn17, Fre18, Hya18}. \\

In those papers, the ``classical'' multivariate tests, which are based on the determinant or the eigenvalues of certain matrices (\textit{e.g.,} Wilks' Lambda, Roy's largest root test, see \cite{Ren, And}), were used. Analogously to their univariate counterparts, they rely on the assumption that the error vectors are i.i.d. multivariate normal, with expectation $\vf{0}$ and identical covariance matrix $\mfg{\Sigma}$ across groups (\textit{i.e.,} homoskedasticity). However, assessing multivariate normality and homoskedasticity is difficult, in particular  when the sample sizes are small. Moreover, even with reasonable sample sizes, there might be good reasons for assuming non-normal errors or heterogeneous covariance matrices. In contrast to the classical multivariate tests discussed in standard textbooks \citep{Ren,Hui}, there are only few methods available that are potentially applicable in a more general setting. Recently, \cite{Fan17} have proposed a rank repeated measures analysis of covariance method; however, only limited simulation evidence regarding the performance in small and possibly unbalanced sample size settings is provided. Apart from that, despite the very broad applicability of the approach of \cite{Fan17}, which renders this method useful for analyzing ordinal data, too, it cannot be regarded as generalizing the scope of classical mean-based MANCOVA
%is not a generalization of the classical MANCOVA in a narrow sense, 
because adjusted mean vectors are not used as effect measures. 
%By contrast, 
As an alternative, the sandwich variance estimation technique
%method 
that was proposed in the context of regression models with clustered errors \citep{Are87} and later examined in combination with various bootstrap procedures \citep{Cam08} stays within a semiparametric framework, but allows for heteroskedasticity.  Although this considerably extends the scope of applications beyond the classical assumptions that were mentioned before, some requirements might still be crucial in a considerable number of practically relevant cases:\\
Firstly, the performance in MANCOVA models with possibly small and unequal group sizes combined with heteroskedasticity and/or non-normality has not been sufficiently examined so far. Secondly, the model under consideration in the cluster-robust estimation approach \citep{Cam08} assumes that the regression parameter vector is one and the same across all subjects and clusters. In the context of regression with clustered errors, this is perfectly fine. However, when being translated to the MANCOVA setting, this means that the model does not allow for regression coefficients, which might vary between the respective coordinates of the outcome vector. But, most likely, the association between a particular covariate and the outcomes might not be the same across all components of the outcome. Thirdly, from a theoretical perspective, it should be noted that the existing methods are based on Wald-type statistics and, therefore, require the covariance matrices to be positive definite. Nevertheless, this assumption might be violated in several ways: 
%For example, in studies on obesity in adolescents, it is of interest to analyze various well-established indices, which are calculated based on ``raw'' laboratory measurements. 
%For example, the HOMA-IR is defined as the product of insulin and glucose levels, scaled by some constant \citep{Mat85, Wal04}. If this index as well as the underlying variables are used as outcomes in a MANCOVA model (\textit{e.g.,} in order to evaluate changes from baseline in different treatment groups), the covariance matrix would obviously be singular. 
%STIMMT NICHT - das muss dann nicht zwingend der Fall sein!! es geht ja nur um Linearkombinationen! - nur dann liegt singularitaet vor!

Especially in psychology, but also in medical research (\textit{e.g.,} quality of life assessments \citep{Hya18, Set18}),  groups of subjects are often compared with respect to a score. Provided that it is sensible to interpret the score as a metric variable -- which requires careful thoughts case-by-case -- employing an ANCOVA model would be appropriate. Most likely, a multivariate ANCOVA would be used, in order to allow for between-group comparisons of, for example, the overall score (\textit{i.e.,} a linear combination of the sub-scores) as well as the subscales on which the score is based.
%on. 
However, the covariance matrix would be singular, then. A similar problem might arise in epilepsy research: Consider the situation of examining whether a particular antiepileptic drug is more efficient than placebo with respect to reducing the number of seizures compared to baseline. In this setting, the change from baseline represents the outcome, and the baseline seizure frequency is regarded as a covariate. However, since it might be sensible to consider not only the overall seizure frequency, but also the reduction in the number of focal and generalized seizures, respectively, we would have a trivariate outcome, and again, the covariance matrix would be singular, then. Apart from singularity due to linear dependencies between the components of the outcome vector, computational issues can also lead to singular covariance structures.    \\

In order to overcome the potential problems mentioned above, we propose a modified ANCOVA ANOVA-type statistic (MANCATS), which is inspired by the modified ANOVA-type statistic proposed for heteroskedastic multivariate analysis of variance without covariates \citep{Fri18}. The main feature of this test statistic is that the ``full'' covariance matrix estimator, which is used in the Wald-type statistic, is replaced by a diagonal matrix that contains only the group-specific variances on the diagonal. So, the ``full'' covariance matrix estimator is not used, yet not being removed completely, as in the ANOVA-type approach (see, for example, \cite{Bru97}). Consequently, our proposed method shares the advantages of both the Wald-type and the ANOVA-type statistic, since the MANCATS is invariant with respect to the scales of the outcome variables, yet not requiring the assumption of positive definite covariance matrices within the groups. Moreover, it should be noted that, in 
%by 
contrast to repeated measures ANCOVA models, the outcomes are allowed to be measured on different scales. In order to approximate the distribution of the MANCATS and improve its finite-sample performance, we apply two different bootstrap procedures. \\

The present manuscript is organized as follows: In Section \ref{Section:GenMancova}, we introduce some notations and set up the model as well as the assumptions 
that are required to ensure the asymptotic validity of the multivariate Wald-type statistic and the two bootstrap MANCATS approaches, respectively. Actually, it turns out that those assumptions are quite weak, which renders our proposed method applicable to a broad range of practically relevant settings. Then, in Section \ref{Section:Bootstrap}, we describe how the two proposed bootstrap procedures work. The finite-sample performance in terms of type I error rates and power is investigated in an extensive simulation study
%, which are 
discussed in Section \ref{Section:Simulations}. In order to further substantiate the findings of the previous section and to illustrate the aforementioned practical applicability, simulations that are based on real-life data from standardized student achievement tests are discussed in Section \ref{Section:RealLife}. Finally, Section \ref{Section:Discussion} contains some closing remarks and ideas for future research. The proofs of the theorems stated in the main body of the manuscript as well as additional simulation results
%, which are referred to in Section \ref{Section:Simulations}, 
are included in the supplementary material.

\section{The general multivariate analysis of covariance (MANCOVA) model}
\label{Section:GenMancova} 
%{\color{red} Konvention: Nur der Index der Kovariable, der von 1 bis c laeuft, steht hochgestellt in Klammern, alle anderen Indizes kommen tiefgestellt vor!}
In the sequel, let $\mf{I}_{m}$ denote the $m$-dimensional identity matrix, and let $\oplus$ and $\otimes$ denote the direct sum and the Kronecker product of matrices, respectively. 
Let $\vf{Y}_{ij} = (Y_{ij1},Y_{ij2},\dots,Y_{ijp})'$ and $\vf{z}_{ij} = (z_{ij}^{(1)},z_{ij}^{(2)},\dots,z_{ij}^{(c)})'$ denote the $p$- and $c$-dimensional outcome and covariate vectors of subject $j$ in group $i$, $i \in \{1,2,\dots,a\}$, $j \in \{1,2,\dots,n_i\}$. We assume that the covariates are fixed, whereas the outcomes are random variables, satisfying the model equation
\begin{equation}
\label{ModelExplicit}
\vf{Y}_{ij} = \vfg{\mu}_{i} + (\vf{z}_{ij}^{\prime}\otimes \ip)\vfg{\nu} + \vfg{\epsilon}_{ij} = \vfg{\mu}_{i} + \sum_{w=1}^{c}z_{ij}^{(w)}\vfg{\nu}^{(w)} + \vfg{\epsilon}_{ij}.
\end{equation}
Thereby, $\vfg{\mu}_{i} = (\mu_{i1},\dots,\mu_{ip})'$ denotes the vector of adjusted means in group $i$, and $\vfg{\nu} = (\vfg{\nu}^{(1)\prime},\dots,\vfg{\nu}^{(c)\prime})'$, where $\vfg{\nu}^{(w)}$ contains the $p$ regression coefficients modelling the association between the $w$-th covariate and the components $1,\dots,p$ of the outcome, $w \in \{1,2,\dots,c\}$. It should be noted that model \refmath{ModelExplicit} allows for unequal regression coefficients for different components of the outcome. However, the coefficients are assumed to be the same across groups, which corresponds to the classical assumption of equal regression slopes in the simple univariate ANCOVA setting. Regarding the errors, we assume that the $\vfg{\epsilon}_{ij}$ are independent, with $E[\vfg{\epsilon}_{ij}] = \vf{0}$ and group-specific covariance matrix $\mfg{\Sigma}_i:= Cov[\vfg{\epsilon}_{ij}]$, $i \in \{1,\dots,a\}, j \in \{1,\dots n_i\}$. \\

Model \refmath{ModelExplicit} can be expressed in a more compact form by using matrix notation: Let $\vf{Y} = (\vf{Y}_{11}^{\prime},\vf{Y}_{12}^{\prime},\dots,\vf{Y}_{an_a}^{\prime})^{\prime}$, $\mf{Z} = (\vf{z}_{11},\vf{z}_{12},\dots,\vf{z}_{an_a})^{\prime}$, $\vfg{\epsilon} = (\vfg{\epsilon}_{11}^{\prime},\vfg{\epsilon}_{12}^{\prime},\dots,\vfg{\epsilon}_{an_a}^{\prime})^{\prime}$, and $\vfg{\mu} = (\vfg{\mu}_1^{\prime},\dots,\vfg{\mu}_a^{\prime})^{\prime}$. Moreover, let $\vf{1}_{m}$ denote the $m$-dimensional vector containing all $1$'s. Then, model \refmath{ModelExplicit} is equivalent to 
\begin{equation}
\label{ModelMnot}
\vf{Y} = \mf{\tilde{M}}\vfg{\mu} + \mf{\tilde{Z}}\vfg{\nu} + \vfg{\epsilon},
\end{equation}
where $\mf{\tilde{M}} = \bigoplus_{i=1}^{a}(\vf{1}_{n_i} \otimes \ip)$, $\mf{\tilde{Z}} = \mf{Z} \otimes \ip$ and $Cov[\vfg{\epsilon}] = \bigoplus_{i=1}^{a}(\mf{I}_{n_i}\otimes \mfg{\Sigma}_i)$. So, in particular, $\vfg{\mu}$ can be estimated by the ordinary least squares estimator $\vfg{\hat{\mu}} = (\vfg{\hat{\mu}}_1^{\prime},\dots,\vfg{\hat{\mu}}_a^{\prime})^{\prime}$, that is,
\[
\vfg{\hat{\mu}}_i = \bar{\vf{Y}}_{i.} - \sum_{w=1}^{c}\bar{z}_{i.}^{(w)}\vfg{\hat{\nu}}^{(w)},
\]
where the dot notation indicates averaging over all subjects in the respective group, and $\vfg{\hat{\nu}}^{(w)}$ denotes the OLS estimator of $\vfg{\nu}^{(w)}$, $w = 1,\dots,c$. We are interested in testing hypotheses about the adjusted mean vectors $\vfg{\mu}_1,\dots,\vfg{\mu}_a$ of the form $H_0: \mf{H}\vfg{\mu} = \vf{0}$, where $\mf{H}$ denotes a contrast matrix (\textit{i.e.,}$\mf{H}\vf{1}_{ap} = \vf{0}$) of full row rank $(a-1)p$. In the present manuscript, however, we specify the hypothesis matrix  in its projector form, that is, we use $\mf{T} := \mf{H}'(\mf{HH}')^{+}\mf{H}$. Thereby, $(\mf{HH}')^{+}$ denotes the Moore-Penrose inverse of $\mf{HH}'$. Note that $\mf{T}$ is uniquely defined, and $ \mf{T}\vfg{\mu} = \vf{0} \Leftrightarrow \mf{H}\vfg{\mu} = \vf{0}$. For example, the hypothesis $H_0: \vfg{\mu}_1 = \ldots = \vfg{\mu}_a$ is equivalent to $H_0: \mf{T}\vfg{\mu} = \vf{0}$, with $\mf{T} = \mf{P}_a \otimes \ip$. Thereby, $\mf{P}_a = \mf{I}_a - \frac{1}{a}\mf{J}_a$, where $\mf{J}_a$ denotes the quadratic $a$-dimensional matrix containing all $1$'s. A more detailed discussion of various multi-factorial and hierarchically nested designs, which are frequently used in practice, can be found in \cite{Kon15}. \\

Now, for testing the hypothesis $H_0:\mf{T}\vfg{\mu} = \vf{0}$, one may consider the Wald-type statistic (WTS)
\begin{equation}
\label{WTS}
W(\mf{T}) = \vfg{\hat{\mu}}^{\prime}\mf{T}^{\prime}(\mf{T}\mfg{\hat{\Sigma}}\mf{T})^{+}\mf{T}\vfg{\hat{\mu}},
\end{equation}
where $\mfg{\hat{\Sigma}}$ denotes the upper-left block of the $2\times 2$ block diagonal matrix $(\mf{\tilde{X}}^{\prime}\mf{\tilde{X}})^{-1}\mf{\tilde{X}}^{\prime}\mf{\hat{S}}\mf{\tilde{X}}(\mf{\tilde{X}}^{\prime}\mf{\tilde{X}})^{-1}$. Thereby, $\mf{\tilde{X}} = (\mf{\tilde{M}},\mf{\tilde{Z}})$, and $\mf{\hat{S}}:= \bigoplus_{i=1}^{a}\bigoplus_{j=1}^{n_i}\mfg{\hat{\Sigma}}_{ij}$ (\textit{i.e.,} $\mf{\hat{S}}$ is a block-diagonal matrix with matrices $\mfg{\hat{\Sigma}}_{11}$,$\mfg{\hat{\Sigma}}_{12}$,$\dots,\mfg{\hat{\Sigma}}_{an_a}$ on the diagonal), and $\mfg{\hat{\Sigma}}_{ij} = \vf{u}_{ij}\vf{u}_{ij}^{\prime}$, $\vf{u}_{ij} = \vf{Y}_{ij} - \vfg{\hat{\mu}}_{i} - (\vf{z}_{ij}^{\prime}\otimes \ip)\vfg{\hat{\nu}}$, $i \in \{1,2,\dots,a\}$, $j \in \{1,2,\dots,n_i\}$. Observe that this covariance matrix estimator basically represents the multivariate generalization of the approach proposed in \cite{Whi} for univariate regression models. Since the publication of this seminal work, several re-scaled estimators have been proposed, with the aim of improving the performance in moderate and small samples \citep{Mac, Cri}. Regarding the theorems stated in the present manuscript, modifications of this type are covered as well, because the scaling factors converge uniformly to $1$. For example, in our simulation studies discussed in Section \ref{Section:Simulations}, we propose a multivariate generalization of the so-called HC4 version of $\mfg{\hat{\Sigma}}_{ij}$, by multiplying the latter with $1/(1-p_{ij,ij})^{\delta_{ij}}$, $\delta_{ij}:= \min\left(4, p_{ij,ij} / (N^{-1}\sum_{r=1}^{a}\sum_{s=1}^{n_r}p_{rs,rs})\right)$, where $p_{rs,rs}$ denotes the diagonal element of the hat matrix $\mf{X}(\mf{X}'\mf{X})^{-1}\mf{X}'$ corresponding to individual $s$ in group $r$, $r \in \{1,2,\dots,a\}, s \in \{1,2,\dots,n_r\}$. Thereby, $\mf{X} = (\mf{M},\mf{Z}) = (\bigoplus_{i=1}^{a}\vf{1}_{n_i}, \mf{Z})$. This adjustment can be interpreted as a natural generalization of the univariate HC4 estimator, which was proposed by Cribari-Neto \citep{Cri} because for $p = 1$, the covariance matrix estimator $\mfg{\hat{\Sigma}}_{ij}$ actually reduces to one single value (\textit{i.e.,} the estimator of the subject-specific variance in the univariate case), and the scaling factor defined above does not depend on the dimension of the outcome. \\

In order to derive the asymptotic distribution of $\mf{W}(\mf{T})$ under $H_0$, the following assumptions are required, where all convergences are understood for $N \to \infty$:
\begin{enumerate}[({M}1)]
\item The errors $\vfg{\epsilon}_{ij}$ are independent, with $E[\vfg{\epsilon}_{ij}] = \vf{0}$, $Cov[\vfg{\epsilon}_{ij}] = \mfg{\Sigma}_i$, and $E(\epsilon_{ijk}^4) \leq C_1 < \infty$, uniformly for all $i \in \{1,2,\dots,a\},j\in \{1,2,\dots,n_i\},k\in\{1,2,\dots,p\}$.  \label{WTSA1}
\item $\mfg{\Sigma}_i > 0$ for all $i \in \{1,2,\dots,a\}$.\label{WTSA2}
\item For all $i \in \{1,2,\dots,a\}: \frac{N}{n_i} \rightarrow \kappa_i > 0$.\label{WTSA3} 
\item The columns of $\mf{Z}$ are linearly independent of each other and of the columns of $\bigoplus_{i=1}^{a}\vf{1}_{n_i}$.\label{WTSA4}
\item For all $i \in \{1,2,\dots,a\}, w \in \{1,\dots,c\}: \frac{1}{n_i}\sum_{j=1}^{n_i}z_{ij}^{(w)} \longrightarrow c_i^{(w)} \in \mathbb{R}$.\label{WTSA5}
\item For all $i \in \{1,2,\dots,a\}: \frac{1}{n_i}\sum_{j=1}^{n_i}\vf{z}_{ij}\vf{z}_{ij}^{\prime} \longrightarrow \mfg{\Xi}_i \in \mathbb{R}^{c\times c}$. \label{WTSA6} 
\end{enumerate}

\noindent
The following theorem establishes the basic result concerning the asymptotic distribution of the Wald-type statistic $W(\mf{T})$:

\begin{thm}
\label{thm1}
Assuming that model \refmath{ModelMnot} as well as \refmathanc{WTSA1} -- \refmathanc{WTSA6} hold, the test statistic $W(\mf{T})$ defined in \refmath{WTS} has an asymptotic ($N\to \infty$) $\chi_f^2$ distribution under $H_0: \mf{T}\vfg{\mu} = \vf{0}$, with $f = rank(\mf{T})$. 
\end{thm}

The assumptions \refmathanc{WTSA1} and \refmathanc{WTSA3} represent standard requirements in an asymptotic framework. Likewise, a violation of \refmathanc{WTSA4} would mean that collinearity was present; in such a case, the validity of the results would be questionable anyway. Conditions \refmathanc{WTSA5} and \refmathanc{WTSA6} are most likely met in virtually any practical application, too, because an unstable average and/or varying spread of the covariates ``on the long run'' (\textit{i.e.,} as more and more subjects are enrolled) would indicate a serious flaw in the design or the conduct of the study. However, \refmathanc{WTSA2} might be violated in a considerable number of practically relevant settings, as outlined in Section \ref{Section:Intro}. This restricts the scope of potential applications of the asymptotic Wald-type test statistic, since it requires positive definiteness. Therefore, we propose a different approach, allowing for possibly singular covariance matrices: Analogously to the modified ANOVA-type statistic (MATS) \citep{Fri18}, we consider a modified ANCOVA ANOVA-type statistic (MANCATS), which is defined as 
\begin{equation}
\label{Mancats}
A(\mf{T}):=  \vfg{\hat{\mu}}^{\prime}\mf{T}^{\prime}(\mf{T}\hat{\mf{D}}\mf{T})^{+}\mf{T}\vfg{\hat{\mu}}.
\end{equation}
Thereby,
\begin{equation}
\label{Dest}
\hat{\mf{D}} = \bigoplus_{i=1}^{a}\frac{1}{n_i}\bigoplus_{k=1}^{p}\hat{\sigma}_{ik}^2 = diag(n_1^{-1}\hat{\sigma}_{11}^2,\dots,n_1^{-1}\hat{\sigma}_{1p}^2,\dots,n_a^{-1}\hat{\sigma}_{ap}^2),
\end{equation}
where 
\begin{equation*}
\hat{\sigma}_{ik}^2 = \frac{1}{n_i - c - 1}\sum_{j=1}^{n_i}u_{ijk}^2,
\end{equation*}
and $u_{ijk} = Y_{ijk} - \hat{\mu}_{ik} - \sum_{w=1}^{c}z_{ij}^{(w)}\hat{\nu}_{k}^{(w)}$ is the residual of outcome $k \in \{1,2,\ldots,p\}$ in group $i \in \{1,2,\ldots,a\}$. It should be noted that if no covariates were included in the model, we would have $c = 0$, resulting in $u_{ijk} = Y_{ijk} - \bar{Y}_{i.k}$. Consequently, $\hat{\sigma}_{ik}^2$ would be equal to the empirical variance estimator, which was used in \cite{Fri18}. So, \refmath{Dest} is a natural generalization of the MATS approach to the MANCOVA setting. As already mentioned before, when using $A(\mf{T})$ instead of $W(\mf{T})$, assumption \refmathm{WTSA2} can be replaced by the weaker requirement 
\begin{enumerate}
\item[(M2)$^{\ast}$] $\mfg{\Sigma}_i \geq 0$ and $\sigma_{ik}^2 > 0$, for all $i \in \{1,2,\dots,a\}$ and $k \in \{1,2,\dots,p\}$.\label{{MANCATSA1}}
\end{enumerate}

This assumption is supposed to be met in a very broad range of practically relevant settings, excluding only cases where, for example, at least one component of the outcome vector is a discrete variable with very few distinct values.\\

Since the ``full'' covariance matrix estimator is replaced by $\hat{\mf{D}}$ in the MANCATS, the asymptotic chi-squared limit distribution from Theorem \ref{thm1} will not hold any more. Nevertheless, there is still at least a formal result regarding the asymptotic distribution of $A(\mf{T})$.  

\begin{thm}
\label{thm2}
Under conditions \refmathanc{WTSA1}, \refmathmancats{{MANCATSA1}}, \refmathanc{WTSA3}--\refmathanc{WTSA6}, and under $H_0: \mf{T}\vfg{\mu} = \vf{0}$, the MANCATS test statistic $A(\mf{T})$ defined in \refmath{Mancats} has, asymptotically ($N \to \infty$), the same distribution as the weighted sum 
\[
U = \sum_{i=1}^{a}\sum_{k=1}^{p}\lambda_{ik}U_{ik}, 
\]
where $U_{ik}\overset{i.i.d.}{\sim}\chi_1^2$, and the weights $\lambda_{ik}$ are the eigenvalues of $\mf{T}(\mf{TDT})^{+}\mf{T}\mfg{\Lambda}_{11}$. Thereby, $\mf{D} = \bigoplus_{i=1}^{a}\bigoplus_{k=1}^{p}\kappa_i\sigma_{ik}^2$, and $\mfg{\Lambda}_{11}$ denotes the upper-left block of the $2\times 2$ block matrix $\mf{\Lambda} = (\mfg{\Xi}^{-1}\otimes\ip)\mfg{\Psi}(\mfg{\Xi}^{-1}\otimes\ip) = \lim_{N \to \infty}[(N^{-1}\mf{\tilde{X}}^{\prime}\mf{\tilde{X}})^{-1}(N^{-1}\mf{\tilde{X}}^{\prime}Cov(\vfg{\epsilon})\mf{\tilde{X}})(N^{-1}\mf{\tilde{X}}^{\prime}\mf{\tilde{X}})^{-1}]$. 
\end{thm}

However, the weights $\lambda_{ik}$ in Theorem \ref{thm2} cannot be calculated, because they represent the eigenvalues of a matrix that contains unknown quantities. Therefore, we will discuss two bootstrap-based approximations in the next section.

\section{Bootstrapping the MANCATS}
\label{Section:Bootstrap}
Firstly, we consider a so-called wild (or multiplier) bootstrap approach, which has been developed and put forward by the work of Wu \citep{Wu}, Liu \citep{Liu} and Mammen \citep{Mam}. Wild bootstrap techniques have already been applied to Wald-type test statistics that are based on White-type covariance matrix estimation techniques in heteroskedastic univariate settings \citep{Cri, Zim18}, in regression models with clustered errors \citep{Cam08} and in multivariate factorial designs \citep{Fri17, Fri18}, in order to improve the small-sample performance of the respective tests. In the present work, we use the following procedure: We generate $N = \sum_{i=1}^{a}n_i$ i.i.d. random variables $T_{ij}$ independently from the data, with $E(T_{11}) = 0$, $Var(T_{11}) = 1$, and $\sup_{i,j}E(T_{ij}^4) < \infty$, $1\leq i \leq a, 1\leq j \leq n_i$. Then, the wild bootstrap observations are defined as 
\[
\vf{Y}_{ij}^{\ast}:= \vf{u}_{ij}\frac{T_{ij}}{\sqrt{1-p_{ij,ij}}}, 1\leq i \leq a, 1\leq j \leq n_i,
\]
where $p_{ij,ij}:=\vf{x}_{ij}^{\prime}(\mf{X}'\mf{X})^{-1}\vf{x}_{ij}$ denotes the diagonal element of the hat matrix corresponding to individual $j$ in group $i$, and $\vf{u}_{ij}$ is the $p$-dimensional residual vector of individual $j$ in group $i$, $i \in \{1,\dots,a\}$, $j \in\{1,\dots,n_i\}$. The scaling factor $(1-p_{ij,ij})^{-1/2}$ has been introduced by \cite{Wu} in case of $p = 1$. Observe that we only generate one single random variable $T_{ij}$ per subject, because coordinate-wise bootstrapping would potentially destroy the dependence structure within subjects. Moreover, it should be mentioned that our proof of Theorem \ref{thm3} works for any particular choice of $T_{ij}$, as long as the fundamental moment conditions mentioned above are met.\\

Once the bootstrap observations have been generated, the bootstrap least squares estimator $\hat{\vfg{\beta}}^{\ast}:=(\mf{\tilde{X}}'\mf{\tilde{X}})^{-1}\mf{\tilde{X}}'\vf{Y}^{\ast}$, the bootstrap covariance matrix estimator $\hat{\mf{D}}^{\ast}:= diag(n_1^{-1}\hat{\sigma}_{11}^{\ast 2},\dots,n_a^{-1}\hat{\sigma}_{ap}^{\ast 2})$ and, by plugging in the bootstrap versions instead of the original estimators in \refmath{Mancats}, the bootstrap analogon $A^{\ast}(\mf{T})$ of the MANCATS test statistic $A(\mf{T})$ are calculated. The following theorem states that using the conditional $(1-\alpha)$-quantile of the empirical distribution of the wild bootstrap test statistic $A^{\ast}(\mf{T})$ as the critical value indeed yields an asymptotically valid test.

\begin{thm}
\label{thm3}
For any parameter vector $\vfg{\beta}:=(\vfg{\mu}^{\prime},\vfg{\nu}^{\prime})^{\prime} \in \mathbb{R}^{(a+c)p}$ and any $\vfg{\beta}_0 = (\vfg{\mu}_0^{\prime},\vfg{\nu}_0^{\prime})' \in \mathbb{R}^{(a+c)p}$ with $\mf{T}\vfg{\mu}_0 = \vf{0}$, we have that given the data, as $N \to \infty$,
\[
\sup_{x\in \mathbb{R}}|P_{\vfg{\beta}}(A^{\ast}(\mf{T})\leq x | \vf{Y}) - P_{\vfg{\beta}_0}(A(\mf{T})\leq x)| \pconv 0
\]
holds, provided that \refmathanc{WTSA1}, \refmathmancats{{MANCATSA1}} and \refmathanc{WTSA3}--\refmathanc{WTSA6} are fulfilled. Thereby, $\pconv$ denotes convergence in probability.
\end{thm}

Secondly, we propose a parametric bootstrap approach, which follows an idea that is similar to existing methods for multivariate analysis of variance models \citep{Kon15, Fri18}. Given the observed data, we impose a group-wise i.i.d. structure in the bootstrap world by drawing
\[
\vf{Y_{ij}^{\star}} \overset{i.i.d.}{\sim} \mathcal{N}(\vf{0},\mfg{\hat{\Sigma}_i}), 1\leq i \leq a, 1\leq j \leq n_i,
\]
where \[
\mfg{\hat{\Sigma}_i}:=\frac{1}{n_i-c-1}\sum_{j=1}^{n_i}\vf{u_{ij}}\vf{u_{ij}^{\prime}}, 1\leq i \leq a.
\]
%So, basically, the group-specific covariance matrices are estimated by using the generalized White approach that has been mentioned in the context of the WTS in Section \ref{Section:GenMancova}. At first, each subject-specific covariance matrix is estimated by the product of the corresponding residual vector times its transpose. Subsequently, these estimates are averaged over all subjects within a group and multiplied with a scaling factor, in order to account for the respective degrees of freedom. 
Once the bootstrap observations have been generated, the parametric bootstrap version $A^{\star}(\mf{T})$ of the MANCATS is obtained analogously to the procedure for the wild bootstrap that was outlined above. Again, the bootstrap test is asymptotically valid:

\begin{thm}
\label{thm4}
For any parameter vector $\vfg{\beta}:=(\vfg{\mu}^{\prime},\vfg{\nu}^{\prime})^{\prime} \in \mathbb{R}^{(a+c)p}$ and any $\vfg{\beta}_0 = (\vfg{\mu}_0^{\prime},\vfg{\nu}_0^{\prime})' \in \mathbb{R}^{(a+c)p}$ with $\mf{T}\vfg{\mu}_0 = \vf{0}$, we have that given the data, as $N \to \infty$,
\[
\sup_{x\in \mathbb{R}}|P_{\vfg{\beta}}(A^{\star}(\mf{T})\leq x | \vf{Y}) - P_{\vfg{\mu}_0}(A(\mf{T})\leq x)| \pconv 0
\]
holds, provided that  \refmathanc{WTSA1}, \refmathmancats{{MANCATSA1}} and \refmathanc{WTSA3}--\refmathanc{WTSA6} are fulfilled.
\end{thm}  

\section{Simulations}
\label{Section:Simulations}

In order to investigate the finite-sample performance of the three methods, which have been discussed in Sections \ref{Section:GenMancova} and \ref{Section:Bootstrap}, we have conducted simulations for a broad range of settings, using \texttt{R} version 3.5.1 \citep{Rco}. In addition to the three test statistics under consideration, we also included the Wilks' Lambda test for comparison.
%as a comparator. 
We chose Rademacher variables as wild bootstrap weights, which means that independently from the data, we generated 
i.i.d.~random variables $T_{11},\ldots,T_{an_a}$ with $P(T_{11} = -1) = P(T_{11} = 1) = 1/2$. These weights have already been applied successfully with respect to the performance of the corresponding tests in the univariate ANCOVA setting \citep{Zim18}. For each scenario, the data generation process was repeated $n_{sim} = 10 000$ times, and within each simulation run, $n_{boot} = 5 000$ bootstrap iterations were performed. We considered data from $a = 2$ groups, with a bivariate outcome (\textit{i.e.,} $p = 2$) and $c = 2$ fixed covariates for each subject. More precisely, the values of the first covariate were equally spaced between $-10$ and $10$, whereas the first and second half of the components of the second covariate vector were equally spaced in $[0,5]$ and $[-2,-1]$, respectively, sorted in descending order. In order to generate the error vectors $\vfg{\epsilon}_{ij}$, at first we drew a random sample of $Np$ independent observations $\xi_{111},\xi_{112},\ldots,\xi_{an_ap}$ from one out of several distributions (normal, $\chi_5^2$, lognormal, double exponential). Then, we transformed them to standardized random variables $\tilde{\xi}_{ijk}:= (\xi_{ijk} - E[\xi_{ijk}]) / (Var[\xi_{ijk}])^{1/2}$ and subsequently calculated the products 
\[
\vfg{\epsilon}_{ij} = \mfg{\Sigma}_i^{1/2}\vfg{\tilde{\xi}}_{ij},
\] 
where $\mfg{\Sigma}_i^{1/2}$ denotes the matrix square root of the covariance matrix of group $i$, and $\vfg{\tilde{\xi}}_{ij}=(\tilde{\xi}_{ij1},\dots,\tilde{\xi}_{ijp})^{\prime}$, $i \in \{1,2,\dots,a\}, j \in \{1,2,\dots,n_i\}$. Doing so, it is ensured that $E[\vfg{\tilde{\xi}}_{ij}] = \vf{0}$ and $Cov[\vfg{\tilde{\xi}}_{ij}] = \mfg{\Sigma}_{i}$. For each of the aforementioned error distributions, we considered three different choices of $\mfg{\Sigma}_1$ and $\mfg{\Sigma}_2$, namely $\mfg{\Sigma}_1 =\mfg{\Sigma}_2 = \mf{I}_2 + 0.5(\mf{J}_2-\mf{I}_2)$ (I), $\mfg{\Sigma}_i = i\cdot\mf{I}_2 + 0.5(\mf{J}_2-\mf{I}_2)$ (II), and $\mfg{\Sigma}_1 = \mfg{\Sigma}_2 = diag(1,0.25) + 0.5(\mf{J}_2-\mf{I}_2)$ (III). Observe that the latter covariance matrix is singular. Finally, the observations were obtained as $\vf{Y}_{ij} = \vfg{\mu}_i + (\vf{z}_{ij}^{\prime}\otimes \ip)\vfg{\nu} + \vfg{\epsilon}_{ij}$, where $\vfg{\nu} = (-0.5,-1.0, 1.5, 1.0)'$. In order to check whether these specifications are sensible, for some scenarios considered in this section, we exemplarily investigated the significance of the conditional associations between the covariates and each of the two outcomes in univariate linear regression models. Exactly speaking, we simulated data for two groups of size $n = 40$ each, assuming lognormal or normal errors, and covariance matrix scenarios I and II. In each of the 4 settings, all $10 000$ simulation runs yielded univariate conditional associations between the components of the outcome and the covariates that were significant at the 5 percent level. Hence, when comparing the means between the groups, an adjustment for covariates is definitely warranted. For the singular scenario III, we set $\vfg{\nu} = (-0.5,-1.0, 1.5, 3.0)'$, in order to reflect the positive linear dependence between the two components of the outcome vector (observe that $(-1.0,3.0) = 2(-0.5,1.5)$). For each of the 12 combinations of distribution and covariance-structure, 4 different sample size scenarios $(n_1,n_2) \in \{(20,20),(10,10),(10,20),(20,10)\}$ were simulated. As mentioned before, for estimating the respective variances and covariances, we multiplied the residual vector of subject $j$ in group $i$ with $(1-p_{ij,ij})^{-\delta_{ij}/2}$. Thereby, $\delta_{ij}:= \min\left(4, p_{ij,ij} / (N^{-1}\sum_{r=1}^{a}\sum_{s=1}^{n_r}p_{rs,rs})\right)$, where $p_{rs,rs}$ denotes the diagonal element of the hat matrix corresponding to individual $s$ in group $r$, $r \in \{1,2,\dots,a\}, s \in \{1,2,\dots,n_r\}$. The adjustment was carried out for all variance and covariance estimators that were used in the three test statistics under consideration. \\

At first, we investigated the empirical type I error rates. Without loss of generality, we set $\vfg{\mu}_i = \vf{0}$, $i\in \{1,2\}$. The results are displayed in Table \ref{Table:TypeIbasic}.  

\begin{table}[h!t]
\centering
\caption{Empirical type I error rates (in \%) of the Wilks' Lambda test (WI), the Wald-type test (WT), the wild bootstrap MANCATS test (MW), and the parametric bootstrap MANCATS test (MP) for $a=2$ groups, $p=2$ dimensions, $c=2$ fixed covariates, and different covariance scenarios, namely I: $\mfg{\Sigma}_i = I_2 + 0.5(J_2-I_2)$, II: $\mfg{\Sigma}_i = i\cdot I_2 + 0.5(J_2-I_2)$, III: $\mfg{\Sigma}_i = diag(1,0.25) + 0.5(J_2-I_2)$, $i \in \{1,2\}$. Errors were drawn either from standard normal, $\chi_5^2$,
%Chi-square(5), 
standard lognormal, or double exponential distribution.}  
\vspace{3mm}
\begin{tabular}{rrc@{\hspace{5pt}}c@{\hspace{5pt}}c@{\hspace{5pt}}c@{\hspace{5pt}}c@{\hspace{5pt}}c@{\hspace{5pt}}c@{\hspace{5pt}}c@{\hspace{5pt}}c@{\hspace{5pt}}c@{\hspace{5pt}}c@{\hspace{5pt}}c@{\hspace{5pt}}c@{\hspace{5pt}}c@{\hspace{5pt}}c@{\hspace{5pt}}c@{\hspace{5pt}}}
\hspace{0.2cm} & \hspace{0.2cm} & \multicolumn{4}{c}{\textbf{Normal}} & \multicolumn{4}{c}{\textbf{Chi-squared(5)}} & \multicolumn{4}{c}{\textbf{Lognormal}}& \multicolumn{4}{c}{\textbf{Double exp.}}\\
$\mf{\Sigma}$& $(n_1,n_2)$ & WI & WT& MW& MP & WI & WT& MW& MP & WI & WT& MW& MP & WI & WT& MW& MP \\
\hline
\multirow{4}{*}{I} & $(20,20)$ & $5.0$ & $8.4$ & $6.6$ & $5.3$ & $5.0$ & $8.2$ & $6.4$ & $4.8$ & $5.0$  & $6.1$ & $5.3$ & $4.5$ & $5.0$ &$7.7$ & $5.8$ & $4.8$ \\
 &$(10,10)$ & $5.3$ & $12.0$ & $7.2$ & $5.0$ & $5.0$ & $11.8$ & $6.7$ & $4.8$& $4.9$ & $8.1$ & $4.6$ & $4.0$ & $4.7$ & $11.2$ & $6.5$ & $4.8$   \\
 &$(10,20)$ & $5.4$ &$12.4$ & $7.0$ & $5.0$ & $4.9$ & $11.6$ & $7.0$ & $5.3$ & $5.4$ &$8.2$ & $5.4$ & $5.0$ & $4.8$ & $11.2$ & $6.9$ & $5.1$ \\
 &$(20,10)$ & $5.2$ &$10.8$ & $7.5$ & $5.1$ & $5.2$& $10.3$ & $7.6$ & $5.1$ &$5.9$&$6.9$ & $5.8$ & $4.4$ & $5.1$ & $10.2$ & $7.0$ & $4.8$ \\
 \hline
\multirow{4}{*}{II} &$(20,20)$ & $2.7$ &$8.1$ & $6.1$ & $5.0$ & $3.0$ & $7.3$ & $6.0$ & $4.6$ & $3.6$ &$5.1$ & $4.8$ & $4.5$ & $2.7$ & $7.1$ & $5.6$ & $4.7$ \\
 & $(10,10)$ & $3.4$ &$11.3$ & $6.7$ & $4.9$ & $3.7$ & $11.0$ & $6.3$ & $4.6$ & $4.1$ & $7.6$ & $4.3$ & $3.9$ & $3.5$ & $10.6$ & $6.2$ & $4.5$ \\
 &$(10,20)$&  $3.7$ &$12.9$ & $7.5$ & $5.2$ &$3.6$& $12.3$ & $7.4$ & $5.2$ & $4.4$&$9.5$ & $5.7$ & $4.9$ & $3.5$ & $11.8$ & $7.3$ & $5.2$ \\
 &$(20,10)$ & $11.5$ &$12.3$ & $7.6$ & $5.0$ & $10.9$ & $12.6$ & $7.4$ & $5.0$ &$10.1$ & $10.1$ & $6.6$ & $4.6$ & $10.9$& $11.9$ & $6.9$ & $4.5$  \\
\hline
\multirow{4}{*}{III} &$(20,20)$ & $-$  &$2.5$&$6.4$ & $5.4$ & $-$& $2.4$ & $6.3$ & $5.2$ &$-$&$1.6$ & $5.7$ & $4.7$ &$-$& $2.1$ & $5.7$ & $4.8$ \\
 &$(10,10)$ &$-$&$3.3$ & $7.4$ & $5.0$ &$-$& $3.0$ & $6.8$ & $4.9$&$-$&$2.2$ & $5.3$ & $4.2$ &$-$& $3.2$ & $6.9$ & $5.0$  \\
 &$(10,20)$ &$-$&$3.4$ & $7.3$ & $4.9$ &$-$& $3.0$ & $7.4$ & $5.1$ &$-$&$2.2$ & $5.5$ & $4.9$ &$-$& $3.1$ & $6.9$ & $5.0$\\
 &$(20,10)$ &$-$&$2.8$ & $7.6$ & $5.0$ &$-$& $2.7$ & $7.6$ & $5.0$ &$-$&$1.7$ & $5.9$ & $4.6$ &$-$& $2.6$ & $6.7$ & $4.5$\\
\hline
\label{Table:TypeIbasic}
\end{tabular}  
\end{table} 

The Wilks' Lambda test showed its well-known conservative behavior in case of positive pairing (\textit{i.e.}, the smallest group has the ``smallest'' variance) and, on the other hand, exceeded the target 5 percent level if negative pairing (\textit{i.e.}, the smallest group has the ``largest'' variance) was present. Although the asymptotic Wald-type test might be considered as a remedy, it tended to be either liberal (covariance settings I and II) or conservative (covariance setting III), even under standard assumptions (\textit{i.e.,} normality and homoskedasticity). As indicated by the results that are displayed in Table S3 in the supplementary material, quite large sample sizes are required in order to achieve accurate type I error control for the Wald-type test. By contrast, the wild bootstrap version of the MANCATS showed a good performance also in small samples, yet being more liberal than the parametric bootstrap MANCATS, which maintained the target level very well in all cases. For neither the asymptotic nor the bootstrap approaches, positive and negative pairing (\textit{i.e.,} setting II with $(n_1,n_2) = (10,20)$ and $(n_1,n_2) = (20,10)$, respectively) affected the results substantially, which could be expected due to the fact that the White estimator and its modified versions have been intended to be ``heteroskedasticity-consistent'' estimators \citep{Whi,Mac10}. However, the results might be dependent on the error distribution, since we see that all approaches yielded slightly smaller type I error rates in the lognormal case compared to the other scenarios. \\

\begin{figure}[h!b]
\centering
\includegraphics[width = 14cm, scale = 1]{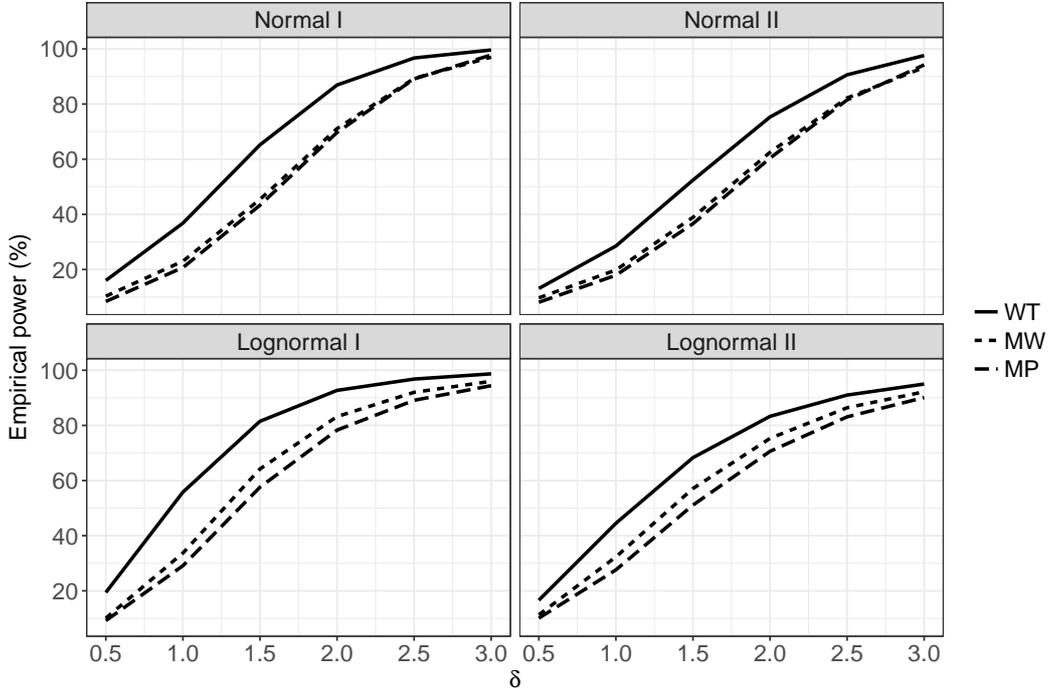}
\caption{Empirical power (in \%) of the Wald-type test (WT), the wild bootstrap MANCATS test (MW), and the parametric bootstrap MANCATS test (MP) for a bivariate outcome ($p=2$) and $c = 2$ covariates. Data were generated for a setting with $a=2$ groups of sizes $n_1 = n_2 = 20$, with $\vfg{\mu}_1=(0,0)'$, $\vfg{\mu}_2 = (\delta,0)'$, and choices of covariance matrices I: $\mfg{\Sigma}_i = I_2 + 0.5(J_2-I_2)$, II: $\mfg{\Sigma}_i = i\cdot I_2 + 0.5(J_2-I_2)$, $i \in \{1,2\}$. Errors were drawn from standard normal or lognormal distribution, respectively.\label{Figure:Power1}}
\end{figure}

\begin{figure}[h!b]
\centering
\includegraphics[width = 14cm, scale = 1]{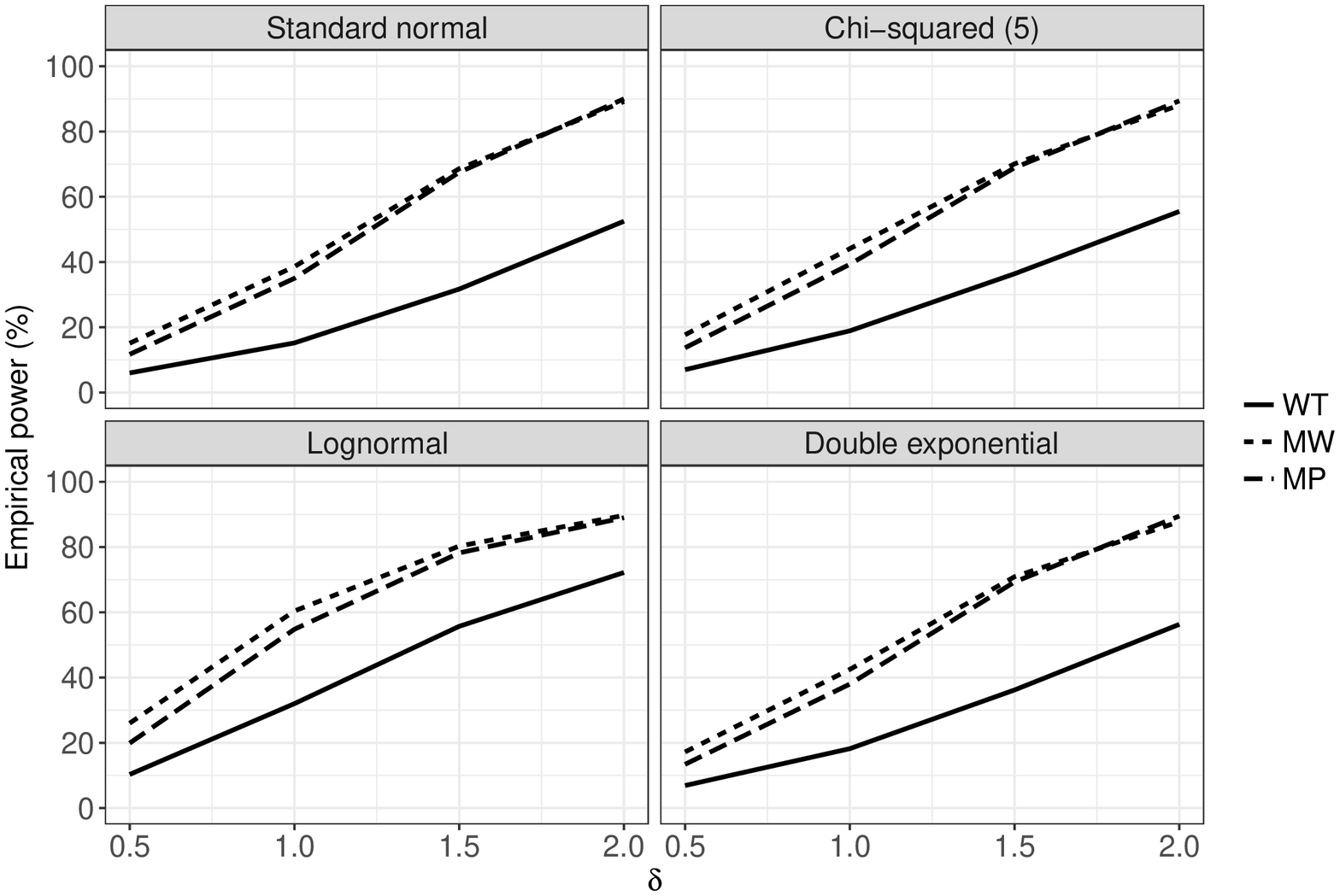}
\caption{Empirical power (in \%) of the Wald-type test (WT), the wild bootstrap MANCATS test (MW), and the parametric bootstrap MANCATS test (MP) for a bivariate outcome ($p=2$) and $c = 2$ covariates. Data were generated for a setting with $a=2$ groups of sizes $n_1 = n_2 = 10$, with $\vfg{\mu}_1=(0,0)'$, $\vfg{\mu}_2 = (\delta,\delta)'$, assuming the singular covariance matrix structure III: $\mfg{\Sigma}_i = diag(1,0.25) + 0.5(J_2-I_2)$, $i \in \{1,2\}$. Errors were drawn from  standard normal, $\chi^2_5$, standard lognormal, or double exponential distribution, respectively.\label{Figure:Power2}}
\end{figure}

Secondly, we simulated the empirical power of the three tests under consideration by staying with the setup that was described at the beginning of this section, but specifying $\vfg{\mu}_1 = (0,0)'$ and $\vfg{\mu}_2 = (\delta,0)'$, with $\delta \in \{0.5,1.0,\dots,3.0\}$. For the singular setting III, we set $\vfg{\mu} = (\delta,\delta)'$, in order to reflect the positive linear dependence between the components of the outcome vector. Although it was difficult to find scenarios where all approaches provided sufficient type I error level control, we tried to cover different distributions and covariance structures. The results are displayed in Figure \ref{Figure:Power1}. Additionally, in order to account for differences in the type I error rates, we provide a plot of the ``achieved power'' (\textit{i.e.}, the difference between empirical power and the scenario-specific type I error rate) in the supplementary material (Figure S1). The two bootstrap MANCATS approaches showed a similar performance, yet having lower power than the Wald-type test. 
%So, obviously, there is some price to pay for the broad scope of applicability and the accurate type I error control of the former.
It should be noted, however, that in case of an underlying normal distribution, the Wald-type approach tended to be liberal (Table \ref{Table:TypeIbasic}). Therefore, the power comparison must be interpreted with caution. Still, in the lognormal case, there was a remarkable difference especially for the homoskedastic setting. However, for the two lognormal scenarios and commonly used target power values around $80$ and $90$ percent (\textit{i.e.}, $\delta \in \{2,2.5\}$), the average power loss of the bootstrap methods compared to the asymptotic Wald-type approach was about $9$ percentage points. The average decrease in achieved power was even lower (about $5$ percentages points). So, summing up, we admit that some caution is needed when applying our proposed bootstrap methods in cases with potentially skewed errors; nevertheless, the price to pay in terms of power might still be acceptable, given the generality of the MANCATS approach. In case of singular covariance matrices, the Wald-type approach should not be used at all, because it had considerably lower power (Figure \ref{Figure:Power2}). In these scenarios, the bootstrap-based tests showed a similar performance, with the wild bootstrap approach being somewhat more powerful, which might be due to its slightly liberal behavior (see Figure S2 in the supplementary material). 

\begin{table}[h!t]

\centering
\caption{Empirical type I error rates (in \%) of the Wilks' Lambda test (WI), the Wald-type test (WT), the wild bootstrap MANCATS test (MW), and the parametric bootstrap MANCATS test (MP) for $a=4$ groups, $p=2$ dimensions and $c=2$ fixed covariates, and different covariance scenarios, namely I: $\mfg{\Sigma}_i = I_2 + 0.5(J_2-I_2)$, II: $\mfg{\Sigma}_i = i\cdot I_2 + 0.5(J_2-I_2)$, III: $\mfg{\Sigma}_i = diag(1,0.25) + 0.5(J_2-I_2)$, $i \in \{1,2,3,4\}$. The group sizes were $\vf{n}_1 = (30,30,30,30)$, $\vf{n}_2 = (15,15,15,15)$, $\vf{n}_3 = (5,10,20,25)$, and $\vf{n}_4 = (25,20,10,5)$. Errors were drawn either from standard normal, $\chi^2_5$, standard lognormal, or double exponential distribution.  }
\vspace{3mm}
\begin{tabular}{rrc@{\hspace{5pt}}c@{\hspace{5pt}}c@{\hspace{5pt}}c@{\hspace{5pt}}c@{\hspace{5pt}}c@{\hspace{5pt}}c@{\hspace{5pt}}c@{\hspace{5pt}}c@{\hspace{5pt}}c@{\hspace{5pt}}c@{\hspace{5pt}}c@{\hspace{5pt}}c@{\hspace{5pt}}c@{\hspace{5pt}}c@{\hspace{5pt}}c@{\hspace{5pt}}}
\hspace{0.1cm} &\hspace{0.4cm} & \multicolumn{4}{c}{\textbf{Normal}} & \multicolumn{4}{c}{\textbf{Chi-squared(5)}} & \multicolumn{4}{c}{\textbf{Lognormal}}& \multicolumn{4}{c}{\textbf{Double exp.}}\\
$\mfg{\Sigma}$ & $N$ & WI & WT& MW& MP & WI & WT& MW& MP & WI & WT& MW& MP & WI & WT& MW& MP \\
\hline
\multirow{4}{*}{I} & $\vf{n}_1$ & $4.2$ & $8.5$ & $5.1$ & $4.3$ & $5.2$ & $8.8$ & $5.2$ & $4.2$ & $4.7$ & $6.5$ &$5.3$ &$4.6$ & $5.0$ & $9.1$ & $5.8$ & $4.9$ \\
&$\vf{n}_2$ & $5.3$ & $13.6$& $6.0$ &$4.6$& $5.1$ & $12.0$& $5.4$& $4.1$& $4.8$ & $8.6$& $5.7$& $4.8$ & $4.9$ & $14.0$& $6.0$& $4.6$\\
&$\vf{n}_3$ & $5.0$ & $13.6$ &$5.7$ &$5.6$ & $4.9$ & $13.8$& $5.6$& $5.8$& $5.1$ & $9.4$& $5.2$& $5.7$ & $5.1$ & $13.0$ & $5.6$ &$5.6$ \\
&$\vf{n}_4$ &$5.0$&$13.7$& $5.3$& $5.6$& $5.2$ & $12.4$& $5.1$& $5.9$& $4.9$ & $8.1$& $4.8$& $6.3$& $4.9$ & $13.0$ & $5.2$ & $5.4$\\
\hline
\multirow{4}{*}{II} & $\vf{n}_1$ & $4.2$ & $8.9$ & $5.0$ & $4.1$ & $4.7$ & $9.3$ & $5.5$ & $4.4$ & $4.4$ & $7.2$ &$5.6$ &$4.6$ & $4.5$ & $9.1$ & $5.7$ & $4.7$ \\
&$\vf{n}_2$ & $4.9$ & $13.9$& $5.8$ &$4.4$&$5.0$& $12.3$& $5.2$& $3.9$& $4.8$ & $9.6$& $5.7$& $5.0$ & $4.6$ & $14.3$& $6.0$& $4.4$\\
&$\vf{n}_3$ & $2.7$ & $13.2$ &$5.4$ &$6.0$ & $2.5$ & $13.2$& $5.5$& $6.2$& $2.6$ & $10.1$& $5.0$& $5.8$ & $2.5$ & $12.5$ & $5.4$ &$5.8$ \\
&$\vf{n}_4$ & $13.5$ &$14.2$& $4.8$& $4.8$& $12.8$ & $12.6$& $4.9$& $5.4$& $10.9$ & $8.5$& $5.4$& $7.1$& $12.6$ & $13.6$ & $4.7$ & $4.8$\\
\hline
\multirow{4}{*}{III} & $\vf{n}_1$ & $-$& $1.1$ & $5.2$ & $4.5$& $-$ & $1.0$ & $5.1$ & $4.3$ & $-$& $0.6$ &$5.0$ &$4.6$ & $-$& $1.1$ & $5.7$ & $5.1$ \\
&$\vf{n}_2$ & $-$& $2.0$& $5.9$ &$4.5$& $-$& $1.7$& $5.3$& $4.3$& $-$& $1.1$& $5.5$& $4.6$ & $-$& $2.0$& $5.7$& $4.6$\\
&$\vf{n}_3$ & $-$& $2.0$ &$5.9$ &$5.6$ & $-$& $1.9$& $5.7$& $5.4$& $-$& $1.0$& $5.2$& $5.8$ & $-$& $1.8$ & $5.5$ &$5.6$ \\
&$\vf{n}_4$ & $-$&$2.1$& $5.3$& $5.4$& $-$& $1.8$& $5.1$& $5.7$& $-$& $0.9$& $4.8$& $5.9$& $-$& $1.7$ & $5.2$ & $5.3$\\
\hline
\label{Table:TypeI4gr2dim}
\end{tabular}  
\end{table}

Furthermore, we conducted several additional simulations, with the aim of investigating whether or not our main findings change substantially under an even broader range of settings. Firstly, all simulation parameters were specified as described above, except from increasing the number of groups to $a=4$, with group size settings $\vf{n}_1 = (30,30,30,30)$, $\vf{n}_2 = (15,15,15,15)$, $\vf{n}_3 = (5,10,20,25)$, and $\vf{n}_4 = (25,20,10,5)$. The results are reported in Table \ref{Table:TypeI4gr2dim}. Compared to the results for the two-group setting, the Wald-type test was even slightly more liberal for covariance matrix scenarios I and II, whereas the wild bootstrap MANCATS yielded type I error rates that were closer to the target 5 percent level. The parametric bootstrap MANCATS again performed well in all scenarios. The most prominent finding was, however, that the conservative behavior of the Wald-type test was even more severe than for $a=2$ groups, with very
%ridiculously 
low empirical type I error rates for some scenarios. By contrast, the MANCATS-based approaches were close to the target 5 percent level. \\

Secondly, we stayed with the two-group setting, but increased the number of dimensions to $p=4$. We modified the covariance matrices accordingly, so, I: $\Sigma_i = I_4 + 0.5(J_4-I_4)$, II: $\Sigma_i = i\cdot I_4+ 0.5(J_4-I_4)$, and III: $\mfg{\Sigma}_i = (\vf{s}_1,\vf{s}_2,\vf{s}_3,\vf{s}_4)$, where $\vf{s}_1 = \vf{s}_2 = \vf{s}_3 = (1,1,1,1/2)'$, $\vf{s}_4 = (1/2,1/2,1/2,1)'$, $i \in \{1,2,3,4\}$. The vectors modelling the associations between a particular covariate and the components of the outcome were set to $\vfg{\nu}^{(1)} = (-0.5,-1.0,-2.0,-0.02)'$ and $\vfg{\nu}^{(2)} = (1.5,1.0,1.0,0.2)'$, respectively. For the singular setting III, we set $\vfg{\nu}^{(1)} = (-0.5,-1.0,-1.5,-0.02)'$ and $\vfg{\nu}^{(2)} = (1.5,3.0,4.5,0.2)'$. Apart from that, all other specifications were the same as before. From Table \ref{Table:TypeI2gr4dim}, it is obvious that in particular the liberality of Wilks' Lambda and the asymptotic Wald-type test was more pronounced than for $p=2$ dimensions. By contrast, the parametric bootstrap MANCATS again yielded type I error rates that were very close to the pre-specified 5 percent level. Moreover, interestingly, the conservative behavior of the asymptotic Wald-type test in case of singular covariance matrices (scenario III) was slightly ameliorated. When further increasing the dimension to $p=8$, the test even became liberal in small and unbalanced group size settings, yet being quite close to the target level for $n_1=n_2 = 20$ (see Table S1 in the supplementary material). These results might indicate that for very small sample sizes, the asymptotic Wald-type test is getting more and more liberal for higher dimensions, but performs reasonably well for moderate sample sizes. Anyway, the parametric bootstrap MANCATS showed very good type I error control regardless of the dimension $p$.\\
Finally, in order to examine whether the results are sensitive to the choice of the covariates, we considered the aforementioned settings for dimension $p=2$ and $a=2$ groups once more, but replaced the values of the second covariate by a random sample from either the standard normal or the standard lognormal distribution, respectively. The results of the simulations for the homoskedastic singular setting III are displayed in Table S2 in the supplementary material. Compared to the corresponding part of Table \ref{Table:TypeIbasic}, the results were similar. The parametric bootstrap MANCATS appeared to perform even slightly better, whereas the asymptotic Wald-type test was somewhat more conservative.

\begin{table}[h!t]

\caption{Empirical type I error rates (in \%) of the Wilks' Lambda test (WI), the Wald-type test (WT), the wild bootstrap MANCATS test (MW), and the parametric bootstrap MANCATS test (MP) for $a=2$ groups, $p=4$ dimensions, $c=2$ fixed covariates, and different covariance scenarios, namely I: $\mfg{\Sigma}_i = I_4 + 0.5(J_4-I_4)$, II: $\mfg{\Sigma}_i = i\cdot I_4 + 0.5(J_4-I_4)$, $i \in \{1,2,3,4\}$. III: $\mfg{\Sigma}_i = (\vf{s}_1,\vf{s}_2,\vf{s}_3,\vf{s}_4)$, where $\vf{s}_1 = \vf{s}_2 = \vf{s}_3 = (1,1,1,1/2)'$, $\vf{s}_4 = (1/2,1/2,1/2,1)'$. Errors were drawn either from standard normal, $\chi^2_5$, standard lognormal, or double exponential distribution.}  
\vspace{3mm}
\begin{tabular}{ccc@{\hspace{5pt}}c@{\hspace{5pt}}c@{\hspace{5pt}}c@{\hspace{5pt}}c@{\hspace{5pt}}c@{\hspace{5pt}}c@{\hspace{5pt}}c@{\hspace{5pt}}c@{\hspace{5pt}}c@{\hspace{5pt}}c@{\hspace{5pt}}c@{\hspace{5pt}}c@{\hspace{5pt}}c@{\hspace{5pt}}c@{\hspace{5pt}}c@{\hspace{5pt}}}
\hspace{0.2cm} & \hspace{0.2cm} & \multicolumn{4}{c}{\textbf{Normal}} & \multicolumn{4}{c}{\textbf{Chi-squared(5)}} & \multicolumn{4}{c}{\textbf{Lognormal}}& \multicolumn{4}{c}{\textbf{Double exp.}}\\
$\mfg{\Sigma}$& $(n_1,n_2)$ & WI& WT& MW& MP &WI& WT& MW& MP & WI& WT& MW& MP & WI&WT& MW& MP \\
\hline
\multirow{4}{*}{I} & $(20,20)$ & $5.1$ & $15.8$ & $6.6$ & $5.2$ &$4.8$ &$14.3$&$6.4$&$5.2$&$4.8$&$11.3$&$5.7$&$4.6$& $5.2$ &$14.7$&$6.5$&$5.2$  \\
 &$(10,10)$ &$5.3$& $28.7$ & $7.2$ & $5.0$ & $5.0$& $27.8$ & $6.5$ & $4.6$ & $4.7$& $23.2$ & $5.1$ & $4.2$ & $5.1$ & $27.9$ & $6.9$ & $4.7$  \\
 &$(10,20)$ &$5.0$&$27.7$ & $7.3$ & $5.2$ &$5.1$&$26.7$&$7.2$&$5.4$&$5.5$&$21.4$&$6.3$&$5.3$&$5.2$&$27.0$&$7.2$&$5.1$\\
 &$(20,10)$ &$5.0$&$27.0$ & $7.6$ & $5.0$ &$5.3$&$24.3$&$7.2$&$4.7$&$6.0$&$22.2$&$6.8$&$4.8$&$5.2$&$25.1$&$7.3$&$4.4$ \\
\hline
\multirow{4}{*}{II} & $(20,20)$&$2.3$&$13.7$&$6.4$&$4.9$&$2.4$&$12.3$&$6.3$&$5.0$&$2.9$&$9.5$&$5.2$&$4.6$&$2.5$&$12.9$&$6.0$&$5.0$ \\
 & $(10,10)$ &$3.7$&$27.0$&$6.6$&$4.8$&$3.6$&$25.9$&$6.0$&$4.3$&$3.9$&$20.7$&$4.5$&$4.1$&$3.5$&$25.7$&$6.3$&$4.5$ \\
 &$(10,20)$&$3.2$ &$28.7$&$7.2$&$5.0$&$3.2$&$28.2$&$7.4$&$5.3$&$4.2$&$21.8$&$5.6$&$4.9$&$3.4$&$28.5$&$7.3$&$5.2$ \\
 &$(20,10)$&$15.1$ &$33.0$&$7.4$&$4.5$&$14.4$ &$30.9$&$6.9$&$4.0$&$12.9$ &$26.6$&$6.5$&$4.3$&$14.9$&$15.1$ &$6.7$&$3.9$ \\
\hline
\multirow{4}{*}{III} &$(20,20)$ &$-$&$2.5$&$6.3$&$5.1$&$-$&$2.5$&$6.4$&$5.0$&$-$&$1.6$&$5.7$&$5.0$&$-$&$2.8$&$6.3$&$5.2$ \\
 &$(10,10)$ &$-$&$5.7$&$7.4$&$5.1$&$-$&$5.2$&$7.1$&$4.7$&$-$&$3.7$&$5.5$&$4.3$&$-$&$5.0$&$6.8$&$4.8$ \\
 &$(10,20)$ &$-$&$5.2$&$7.4$&$4.9$&$-$&$5.2$&$7.4$&$5.3$&$-$&$3.0$&$5.8$&$5.0$&$-$&$5.0$&$7.3$&$5.0$ \\
 &$(20,10)$ &$-$&$4.6$&$7.6$&$5.0$&$-$&$4.0$&$7.2$&$4.7$&$-$&$2.8$&$6.4$&$4.5$&$-$&$4.0$&$7.0$&$4.8$ \\
\hline
\label{Table:TypeI2gr4dim}
\end{tabular}  
\end{table}

\clearpage

\section{Real-life data example}
\label{Section:RealLife}

In this section, we consider standardized test data, which has been collected by William D. Rohwer (University of California, Berkeley). The dataset was taken from \cite{Tim}. For each of the $N = 69$ kindergarten children
%students 
that were included in the study, the socioeconomic status (low/high), the number of points on three different achievement tests, and the performance on 5 paired associate learning tasks were recorded. For example, it might be of interest to compare the results from the achievement tests between the two socioeconomic groups, while adjusting for the performance on the learning tasks. For ease of illustration, we shall consider a two-group bivariate MANCOVA model with the performance on the Peabody Picture Vocabulary Test (PPVT) and the Student Achievement Test (SAT) as the components of the outcome vector, and the sum of the 5 paired associate learning task results as a covariate. At first, calculating the estimated residual covariance matrices for the high- $(n=32)$ and low-socioeconomic-status groups ($n = 37$) yielded 
\[
\hat{\mfg{\Sigma}}_h^{(r)} = \left(\begin{matrix}
145.88 & 113.18 \\
113.18 & 1073.21
\end{matrix}\right)\,\,\text{and}\,\, \hat{\mfg{\Sigma}}_{\ell}^{(r)} = \left(\begin{matrix}
99.07 & 60.85 \\
60.85 & 458.41
\end{matrix}\right),
\]
respectively. Secondly, we considered the following practically relevant setting: In general, when comparing test results between groups, the primary research question is often focused on differences in a particular overall score. Of course, however, the comparison of subscales might also be of interest. Ideally, both the overall score as well as the subscales, which make up the former overall value, should be analyzed jointly (\textit{i.e.}, by merging them into one single vector and subsequently applying a multivariate method), since the correlation would then be accounted for. By contrast, if two separate tests were applied, the correlation would be ignored. It should be noted that such situations might also arise quite frequently when analyzing, for example, global assessment variables in medical research \citep{IchE9}. However, in cases where the total score is a linear combination (\textit{e.g.}, the sum) of the subscores, the group-specific covariance matrices will be singular. We shall consider such a scenario in the sequel, by adding the sum of the two achievement test results as a third component to the outcome vector, while leaving all other specifications unchanged. The corresponding group-specific estimated residual covariance matrices were
\[
\hat{\mfg{\Sigma}}_h^{(s)} = \left(\begin{matrix}
145.88 & 113.18 & 259.06 \\
113.18 & 1073.21 & 1186.39\\
259.06 & 1186.39 & 1445.45
\end{matrix}\right)\,\,\text{and}\,\, \hat{\mfg{\Sigma}}_{\ell}^{(s)} = \left(\begin{matrix}
99.07 & 60.85 & 159.92 \\
60.85 & 458.41 & 519.26\\
159.92 & 519.26 & 679.19
\end{matrix}\right),
\]
respectively. Observe that heteroskedasticity is present both in the regular and in the singular setting. In order to investigate whether or not the performance of the approaches that are considered in the present manuscript was similar to the results from Section \ref{Section:Simulations}, we examined the empirical type I error rates, adopting the same specificiations for the data generation process as in the previous section, but with $\mfg{\Sigma}_1$ and $\mfg{\Sigma}_2$ replaced by $\hat{\mfg{\Sigma}}_h^{(r)}$ and $\hat{\mfg{\Sigma}}_l^{(r)}$, or $\hat{\mfg{\Sigma}}_h^{(s)}$ and $\hat{\mfg{\Sigma}}_l^{(s)}$, respectively. Regarding the group sizes, apart from the original values (\textit{i.e.,} $(n_1,n_2) = (32,37)$), we also considered two additional scenarios, namely $(n_1,n_2) = (23,46)$ and $(n_1,n_2) = (46,23)$ (note that if we were in a randomized clinical trial setting, these group sizes would correspond to 1:2 and 2:1 allocation ratios). These specifications allow for investigating the effects of positive and negative pairing, while leaving the total sample size $N = 69$ unchanged. As in Section \ref{Section:Simulations}, the target alpha level was set to 5 percent, and the number of simulation and bootstrap runs were $n_{sim} = 10000$ and $n_{boot} = 5000$, respectively. The results are displayed in Table \ref{Table:TypeITimm}. Obviously, the empirical type I error rates of Wilks' Lambda deviated from the target level in presence of negative and positive pairing. By contrast, especially the parametric bootstrap MANCATS performed very well. Interestingly, the asymptotic Wald-type test showed a less liberal behavior compared to the settings in Section \ref{Section:Simulations}, which might be explained by the fact that the total sample size $N = 69$ was larger. In the singular case, however, the Wald-type test was conservative again, whereas the two bootstrap-based approaches were close to the target level. Only in case of positive pairing and lognormal errors, our proposed MANCATS tests tended to yield liberal results. This  might be explained by the somewhat suboptimal behavior of the White estimator in terms of bias and variance, which has been observed in the univariate case \citep{Zim18}. Summing up, apart from some issues in that particular scenario, the parametric bootstrap MANCATS was close to the target type I error level in all settings and might therefore be considered as the method of choice for conducting MANCOVA analyses of small to moderately large samples.

\begin{table}[h!t]
\centering
\caption{Empirical type I error rates (in \%) of the Wilks' Lambda test (WI), the Wald-type test (WT), the wild bootstrap MANCATS test (MW), and the parametric bootstrap MANCATS test (MP) for $a=2$ groups (low vs. high), $c=1$ covariate (sum of the learning task results), and the following covariance matrix settings: Regular scenario (R), where $\mfg{\Sigma}_1 = \hat{\mfg{\Sigma}}_h^{(r)}$ and $\mfg{\Sigma}_2 = \hat{\mfg{\Sigma}}_{\ell}^{(r)}$; singular scenario (S), with $\mfg{\Sigma}_1 = \hat{\mfg{\Sigma}}_h^{(s)}$ and $\mfg{\Sigma}_2 = \hat{\mfg{\Sigma}}_{\ell}^{(s)}$. Errors were drawn either from standard normal, $\chi^2_5$, standard lognormal, or double exponential distribution.}  
\vspace{3mm}
\begin{tabular}{rrc@{\hspace{5pt}}c@{\hspace{5pt}}c@{\hspace{5pt}}c@{\hspace{5pt}}c@{\hspace{5pt}}c@{\hspace{5pt}}c@{\hspace{5pt}}c@{\hspace{5pt}}c@{\hspace{5pt}}c@{\hspace{5pt}}c@{\hspace{5pt}}c@{\hspace{5pt}}c@{\hspace{5pt}}c@{\hspace{5pt}}c@{\hspace{5pt}}c@{\hspace{5pt}}}
\hspace{0.2cm} & \hspace{0.2cm} & \multicolumn{4}{c}{\textbf{Normal}} & \multicolumn{4}{c}{\textbf{Chi-square(5)}} & \multicolumn{4}{c}{\textbf{Lognormal}}& \multicolumn{4}{c}{\textbf{Double exp.}}\\
$\mf{\Sigma}$& $(n_1,n_2)$ & WI & WT& MW& MP & WI & WT& MW& MP & WI & WT& MW& MP & WI & WT& MW& MP \\
\hline
\multirow{3}{*}{R} & $(32,37)$ & $5.9$ & $5.9$ & $5.1$ & $4.9$ & $6.0$ & $6.1$ & $5.1$ & $5.0$ & $5.6$  & $6.2$ & $6.0$ & $5.0$ & $5.7$ &$6.2$ & $5.5$ & $5.3$ \\
 &$(23,46)$ & $8.4$ & $6.7$ & $5.5$ & $5.3$ & $9.0$ & $7.0$ & $5.6$ & $5.2$& $7.5$ & $9.8$ & $9.2$ & $8.0$ & $8.2$ & $7.0$ & $5.7$ & $5.6$   \\
 &$(46,23)$ & $2.8$ &$5.7$ & $5.1$ & $5.0$ & $2.6$ & $5.7$ & $5.1$ & $4.9$ & $3.4$ &$4.4$ & $4.9$ & $4.0$ & $3.2$ & $6.0$ & $5.1$ & $5.1$ \\
 \hline
\multirow{3}{*}{S} &$(32,37)$ & $-$ &$2.5$ & $5.1$ & $4.9$ & $-$ & $2.7$ & $5.5$ & $5.4$ & $-$ &$2.5$ & $5.4$ & $4.7$ & $-$ & $2.8$ & $5.2$ & $5.0$ \\
 & $(23,46)$ & $-$ &$3.0$ & $5.1$ & $5.1$ & $-$ & $3.7$ & $5.6$ & $5.3$ & $-$ & $4.2$ & $7.4$ & $6.7$ & $-$ & $3.1$ & $5.7$ & $5.6$ \\
 &$(46,23)$&  $-$ &$2.6$ & $5.3$ & $5.1$ &$-$& $2.5$ & $4.8$ & $4.7$ & $-$&$1.9$ & $5.0$ & $4.3$ & $-$ & $2.3$ & $5.1$ & $4.9$ \\
\hline
\label{Table:TypeITimm}
\end{tabular}  
\end{table} 

\section{Discussion and conclusions}
\label{Section:Discussion}

In the present manuscript, we have proposed an alternative to the MANCOVA Wald-type test statistic. Although the latter already allowed for heteroskedasticity by incorporating a multivariate generalization of the White sandwich estimator, 
it still required assumptions which are not met
%it might still require assumptions, which might not be met 
in a considerable number of practically relevant settings. In particular, if the covariance matrix is singular due to computational reasons or linear dependencies between the components of the outcome vector, it is not appropriate to use the Wald-type statistic. Therefore, we have proposed and examined a modified ANOVA-type statistic for the MANCOVA setting, where the full covariance matrix estimator is replaced by a diagonal matrix, which contains basically the re-scaled group-wise variance estimators of the outcomes on the diagonal. As a consequence, the positive definiteness of the group covariance matrices is not required any more. We have proposed two different bootstrap versions of the MANCATS, which both yield an asymptotically valid test. We would like to emphasize that the respective proofs do not require any additional assumptions compared to the Wald-type setting. Moreover, the results of extensive simulations show that the MANCATS-based approaches outperform the asymptotic Wald-type test statistic in most scenarios under consideration, for singular as well as non-singular covariance matrix specifications. On top of that, the Wald-type test statistic is clearly inferior in terms of power in singular settings and, therefore, should not be used. So, summing up, both MANCATS tests can be recommended and are preferable compared to the classical asymptotic Wald-type test statistic, except perhaps for the case of lognormal (\textit{i.e.,} skewed) errors, where all three methods have a somewhat suboptimal performance.    \\

There is a wide variety of different bootstrap methods available; the same applies to heteroskedasticity-consistent covariance matrix estimation techniques. Hence, not all combinations of methods from these fields can be considered. Nevertheless, we have carefully thought about some approaches, which are not covered in the present manuscript. Firstly, various variants of White's sandwich estimator \citep{Whi} for uni- and multivariate linear models have been proposed in the literature (see, for example, \cite{Mac10} for an overview). We decided to use a multivariate generalization of the HC4 estimator \citep{Cri}, since its performance turned out to be superior to other variants (HC0, HC2) in the univariate case \citep{Zim18}. Secondly, one might wonder why a nonparametric bootstrap procedure was not considered, given that it is straightforward to implement, and it has been used in multivariate factorial designs \citep{Kon15,Fri18}. The reason is that in the context of heteroskedastic regression models, the nonparametric bootstrap has suboptimal properties with respect to variance estimation \citep{Wu}. Moreover, in multivariate factorial designs, the nonparametic bootstrap performed somewhat worse than the respective competitors \citep{Kon15, Fri18}.\\

We would like to emphasize that all three methods, which were discussed in our manuscript, are applicable to a broad range of practically relevant settings not only due to the generality of the underlying assumptions (\textit{e.g.,} allowing for group-specific covariance matrices, and not assuming any particular distribution of the errors), but also with respect to the model we have considered. In particular, the associations between the covariates and a certain component of the outcome vector are allowed to vary from coordinate to coordinate. Nevertheless, still, the scope of applications is restricted to metric outcome variables. Some evidence indicates that with respect to type I errors and power, ANCOVA methods might be valid even for ordinal outcomes \citep{Sul03}. However, this cannot be regarded as a satisfactory solution, both from a theoretical point of view, and with respect to the interpretation of the estimates. Since the case of ordinal outcomes is highly relevant in practice, as briefly outlined in Section \ref{Section:Intro}, efforts towards developing methods which allow for MANCOVA-like analyses of multivariate ordinal outcomes should be made in future research. 

\iffalse
%weiß nicht, ob man das so genau schreiben sollte....
To this end, applying probabilistic index models (PIM) might be a promising approach \citep{Tha,DeN15,Amo18}. Apart from that, the authors conjecture that the ideas discussed in the present manuscript might be useful for ANCOVA models with clustered errors, too. Maybe, similarly to the MANCATS, it would suffice to apply a cluster-wise wild bootstrap procedure, even without subsequently using a cluster-robust covariance matrix estimator \citep{Cam08}. This would substantially facilitate calculations and most likely reduce the computational burden. Therefore, this topic might also be a promising goal of future research.           \\
\fi

\clearpage

% change layout of figure and table labels
\setcounter{table}{0}
\setcounter{figure}{0}
\setcounter{section}{0}
\setcounter{equation}{0}
\renewcommand{\thetable}{S\arabic{table}}  
\renewcommand{\thefigure}{S\arabic{figure}} 

\noindent

{\Huge Supplementary Material}

\vspace{3cm}

Section \ref{Supp:proofs} contains the proofs of Theorems 1 to 3, which are stated in the manuscript. Some additional simulation results concerning the type I error rates in settings with higher dimension, different choices of the covariates, and large sample sizes are provided in Section \ref{Supp:typeI}. Finally, in order to account for differences in type I error rates when comparing the MANCATS and the White approaches with respect to empirical power, we display achieved power values (\textit{i.e.,} empirical power minus the corresponding type I error rate) in some plots in Section \ref{Supp:achieved}.

\section{Proofs}
\label{Supp:proofs}

\subsection{Notations and assumptions}
\label{proofass}
Let $\mf{I}_{m}$ denote the $m$-dimensional identity matrix, and let $\oplus$ and $\otimes$ denote the direct sum and the Kronecker product of matrices, respectively.
In order to simplify notations, we consider $\vf{Y} = (\xm \otimes \ip) \bpar + \vfg{\epsilon}$, where $E(\vfg{\epsilon})= \vf{0}, \mf{S}:= Cov(\vfg{\epsilon}) = \bigoplus_{i=1}^{a}(\mf{I_{n_i}}\otimes \mfg{\Sigma}_i)$, $\xm = (\bigoplus_{i=1}^{a}\vf{1}_{n_i}, \mf{Z})$, $\bpar = (\vfg{\mu}',\vfg{\nu}')'$. Moreover, let $\best = (\vfg{\hat{\mu}}',\vfg{\hat{\nu}}')'$ denote the ordinary least squares estimator of $\bpar$, $\SigmaFull = ((\xm'\xm/N)^{-1}\otimes\ip)(N^{-1}(\xm^{\prime}\otimes\ip)\mf{S}(\xm\otimes\ip))((\xm'\xm/N)^{-1}\otimes\ip)$, $\SigmaWhiteFull = ((\xm'\xm/N)^{-1}\otimes\ip)(N^{-1}(\xm^{\prime}\otimes\ip)\mf{\hat{S}}(\xm\otimes\ip))((\xm'\xm/N)^{-1}\otimes\ip)$, $\mf{\hat{S}} = \bigoplus_{i=1}^{a}\bigoplus_{j=1}^{n_i}\mfg{\hat{\Sigma}}_{ij}$, 
$\mfg{\hat{\Sigma}}_{ij} = \vf{u}_{ij}\vf{u}_{ij}^{\prime}$, $N = \sum_{i=1}^{a}n_i$. Thereby, $\vf{u}_{ij}$ denotes the $p$-dimensional residual vector corresponding to subject $j$ in group $i$, $1\leq i \leq a$, $1\leq j \leq n_i$. \\
Observe that assumption (M5) implies that the elements of $\xm$ are uniformly bounded (at least for sufficiently large $N$). Moreover, $(\frac{1}{N}\xm'\xm)$ converges to, say, $\mfg{\Xi}$, due to (M3), (M5), and (M6). According to condition (M4), $\xm$ is of full column rank, which in turn implies that $\frac{1}{N}\xm'\xm$ is positive definite. Consequently, $(\frac{1}{N}\xm'\xm)^{-1}$ converges to $\mfg{\Xi}^{-1}$. Analogously, we get that $\frac{1}{N}(\xm'\otimes\ip)\mf{S}(\xm\otimes\ip)$ converges to a matrix $\Psi$, which is positive definite if we additionally assume (M2). In the following proofs, we omit all scaling factors that are based on elements of the hat matrix, for sake of notational simplicity. Observe that the derivations are still valid, though, because $|p_{ij}| = |\vf{x}_{ij}'(\mf{X}'\mf{X})^{-1}\vf{x}_{ij}| \leq C/N$. 

\vspace{1cm}

\subsection{Proof of Theorem 1}
\label{proof1}
The proof comprises two main steps:
{\renewcommand{\theenumi}{(A\arabic{enumi})}
\renewcommand{\labelenumi}{\theenumi}
\begin{enumerate}

\item $\sqrt{N}(\vfg{\hat{\beta}}-\vfg{\beta}) \dconv \mathcal{N}(\vf{0},\mfg{\Lambda})$, where $\mfg{\Lambda}:=(\mfg{\Xi}^{-1}\otimes \ip)\mfg{\Psi}(\mfg{\Xi}^{-1}\otimes\ip)$, $\mfg{\Psi}:= \lim_{N\to \infty}(N^{-1}(\xm'\otimes\ip)\mf{S}(\xm\otimes \ip))$, $\mfg{\Xi}^{-1}:= \lim_{N \to \infty}(N^{-1}\xm'\xm)^{-1}$. Note that both limits are well-defined (cf. Section \ref{proofass}).\label{wts1}
\item $\SigmaWhiteFull - \SigmaFull \pconv \mf{0}$. \label{wts2}
\end{enumerate}
}

\noindent
\textit{Proof of \ref{wts1}:}\\[5mm]
Let $\xm = (\vf{x}_{11},\vf{x}_{12},\dots,\vf{x}_{an_a})^{\prime}$. Doing some algebra yields 
\begin{equation*}
\sqrt{N}(\best - \bpar) = \sqrt{N}((\xm^{\prime}\xm)^{-1}\xm^{\prime}\otimes \ip)\vfg{\epsilon} = \sqrt{N}((\xm^{\prime}\xm)^{-1}\otimes \ip)\sum_{i=1}^{a}\sum_{j=1}^{n_i}(\vf{x}_{ij}\otimes \ip)\vfg{\epsilon}_{ij}.
\end{equation*}
Hence, we define the triangular array $\vfg{\eta}_{ij}:= \sqrt{N}((\xm^{\prime}\xm)^{-1}\otimes \ip)(\vf{x}_{ij}\otimes \ip)\vfg{\epsilon}_{ij}, 1\leq i \leq a$, $1\leq j \leq n_i$, and apply the multivariate Lindeberg-Feller theorem.
Obviously, $E(\vfg{\eta}_{ij}) = \vf{0}$, and $(\vfg{\eta}_{ij})_{i,j}$ are independent, due to (M1). Secondly, we get
\begin{align}
\sum_{i=1}^{a}\sum_{j=1}^{n_i}Cov(\vfg{\eta}_{ij}) &= \sum_{i=1}^{a}\sum_{j=1}^{n_i} N ((\xm^{\prime}\xm)^{-1}\vf{x}_{ij}\otimes \ip)Cov(\vfg{\epsilon}_{ij})(\vf{x}_{ij}^{\prime}(\xm^{\prime}\xm)^{-1}\otimes \ip)\nonumber\\
&= N\sum_{i=1}^{a}\sum_{j=1}^{n_i} ((\xm^{\prime}\xm)^{-1}\vf{x}_{ij}\otimes \ip)\mfg{\Sigma}_i(\vf{x}_{ij}^{\prime}(\xm^{\prime}\xm)^{-1}\otimes \ip)\nonumber\\
&= (\xm^{\prime}\xm/N)^{-1}\otimes\ip)(N^{-1}(\xm'\otimes\ip)\mf{S}(\xm\otimes \ip))(\xm^{\prime}\xm/N)^{-1}\otimes\ip)\nonumber\\
&\rightarrow \mfg{\Lambda}.\nonumber
\end{align}
Finally, we have to show that the Lindeberg condition
holds. That is, 
\begin{equation}
\sum_{i=1}^{a}\sum_{j=1}^{n_i}E(||\vfg{\eta}_{ij}||^2 \mathbbm{1}\{||\vfg{\eta}_{ij}||^2 > \delta\}) \rightarrow 0\, \forall \delta > 0.
\label{LindebergCond}
\end{equation}

Firstly, 
\begin{align}
||\vfg{\eta}_{ij}||^2 &= \vfg{\eta}_{ij}^{\prime}\vfg{\eta}_{ij} = N \vfg{\epsilon}_{ij}^{\prime}(\vf{x}_{ij}^{\prime}(\xm^{\prime}\xm)^{-1}\otimes\ip) ((\xm^{\prime}\xm)^{-1}\vf{x}_{ij}\otimes\ip)\vfg{\epsilon}_{ij}\nonumber\\
&= \frac{1}{N} \vfg{\epsilon}_{ij}^{\prime}(\vf{x}_{ij}^{\prime}(\xm^{\prime}\xm/N)^{-1}\otimes\ip) ((\xm^{\prime}\xm/N)^{-1}\vf{x}_{ij}\otimes\ip)\vfg{\epsilon}_{ij}\nonumber.\label{wtsnorm}
\end{align}

Let $\vfg{\tilde{\eta}}_{ij}:= ((\xm^{\prime}\xm/N)^{-1}\otimes\ip)(\vf{x}_{ij}\otimes \ip)\vfg{\epsilon}_{ij}$ (consequently, $||\vfg{\eta}_{ij}||^2 = N^{-1} ||\vfg{\tilde{\eta}}_{ij}||^2$). %Firstly, recall that $
%\SigmaFull = ((\xm^{\prime}\xm/N)^{-1}\otimes \ip)\left(\frac{1}{N}%(\xm^{\prime}\otimes \ip) S(\xm\otimes \ip)\right)((\xm^{\prime}\xm/N)^{-1}%\otimes \ip)$
%Note that our assumptions ensure that $\SigmaFull$ is well-defined and %invertible (cf. Section \ref{proofass}), and $\SigmaFull^{-1} = %((\xm^{\prime}\xm/N)\otimes \ip)(N^{-1}(\xm^{\prime}\otimes \ip) S(\xm\otimes %\ip))^{-1}((\xm^{\prime}\xm/N)\otimes \ip)$.
%Using Lancaster's theorem, we get
%\begin{align*}
%E(||\vfg{\tilde{\eta}}_{ij}||^2) &=tr\left((\vf{x}_{ij}^{\prime}\otimes \ip) \left[\frac{1}{N}(\xm'\otimes\ip)\mf{S}(\xm\otimes\ip)\right]^{-1}(\vf{x}_{ij}\otimes \ip) \mfg{\Sigma}_i\right)\\
%&= tr\left((\vf{x}_{ij}\otimes \ip)\mfg{\Sigma}_i(\vf{x}_{ij}^{\prime}\otimes \ip) \left[\frac{1}{N}(\xm'\otimes\ip)\mf{S}(\xm\otimes\ip)\right]^{-1}\right).
%\end{align*}
Using Lancaster's theorem, we get
\begin{align*}
E(||\vfg{\tilde{\eta}}_{ij}||^2) &=tr\left((\vf{x}_{ij}^{\prime}\otimes \ip) \left[\frac{1}{N}(\xm'\xm\otimes\ip)\right]^{-1}\left[\frac{1}{N}(\xm'\xm\otimes\ip)\right]^{-1}(\vf{x}_{ij}^{\prime}\otimes \ip) \mfg{\Sigma}_i\right)\\
&= tr\left((\vf{x}_{ij}\otimes \ip)\mfg{\Sigma}_i(\vf{x}_{ij}^{\prime}\otimes \ip) \left[\frac{1}{N}(\xm'\xm\otimes\ip)\right]^{-1}\left[\frac{1}{N}(\xm'\xm\otimes\ip)\right]^{-1}\right).
\end{align*}

So, we have to calculate the trace of a product of matrices, which are uniformly bounded element-wise and have dimensions independent of $N$. Therefore, 
\begin{equation}
\sup_{i,j,N}E(||\vfg{\tilde{\eta}}_{ij}||^2) \leq C\,\,\text{for some } C>0.\label{LindeExp}
\end{equation}

Next, let $\mf{A}_{ij}:=(\vf{x}_{ij}^{\prime}\otimes \ip)((\xm'\xm/N)^{-1}\otimes \ip)((\xm'\xm/N)^{-1}\otimes \ip)(\vf{x}_{ij}\otimes \ip)$. Then, $||\vfg{\tilde{\eta}}_{ij}||^2$ simplifies to $\vfg{\epsilon}_{ij}^{\prime}\mf{A}_{ij}\vfg{\epsilon}_{ij}$, which can be explicitely written as
\[
||\vfg{\tilde{\eta}}_{ij}||^2 = \vfg{\epsilon}_{ij}^{\prime}\mf{A}_{ij}\vfg{\epsilon}_{ij} = \sum_{k=1}^{p}\sum_{\ell = 1}^{p}a_{ij,k\ell}\epsilon_{ijk}\epsilon_{ij\ell},
\]
where $a_{ij,k\ell}$ denotes the element in the $k$-th row and the $\ell$-th column of $\mf{A}_{ij}$. Consequently, the expectation $E[||\vfg{\tilde{\eta}}_{ij}||^4]$ is equal to a sum of $p^4$ expectations of products $\epsilon_{ij{k_1}}\epsilon_{ij{\ell_1}}\epsilon_{ij{k_2}}\epsilon_{ij{\ell_2}}$, scaled with elements of the matrix $\mf{A_{ij}}$. The latter quantities are uniformly bounded from above, and using the Cauchy-Schwarz inequality iteratively, the expectation of the product of the error terms can be bounded from above by $E[\epsilon_{ij{k_1}}^4]E[\epsilon_{ij{\ell_1}}^4]E[\epsilon_{ij{k_2}}^4]E[\epsilon_{ij{\ell_2}}^4]$, which is uniformly bounded, due to assumption (M1). So, all in all, the fourth moments $E[||\vfg{\tilde{\eta}}_{ij}||^4]$ can be uniformly bounded from above by, say, $\tilde{C}$. Consequently,
\begin{align*}
\sum_{i=1}^{a}\sum_{j=1}^{n_i}E[||\vfg{\eta}_{ij}||^2 \mathbbm{1}\{||\vfg{\eta}_{ij}||^2 > \delta\}] &= 
\frac{1}{N}\sum_{i=1}^{a}\sum_{j=1}^{n_i}E[||\vfg{\tilde{\eta}}_{ij}||^2 \mathbbm{1}\{||\vfg{\tilde{\eta}}_{ij}||^2 > \delta N\}] \\
&\leq \frac{1}{N}\sum_{i=1}^{a}\sum_{j=1}^{n_i}\sqrt{E[||\vfg{\tilde{\eta}}_{ij}||^4]}\sqrt{E[\mathbbm{1}\{||\vfg{\tilde{\eta}}_{ij}||^2 > \delta N\}]}\\
&\leq \frac{\tilde{C}}{N}\sum_{i=1}^{a}\sum_{j=1}^{n_i}\sqrt{E[\mathbbm{1}\{||\vfg{\tilde{\eta}}_{ij}||^2 > \delta N\}]}\\
&\leq\tilde{C}\left(\frac{C}{\delta N}\right)^{1/2},
\end{align*} 
where the last step was due to 
\begin{align*}
E[\mathbbm{1}\{||\vfg{\tilde{\eta}}_{ij}||^2 > \delta N\}] &= P(||\vfg{\tilde{\eta}}_{ij}||^2 > \delta N)\leq \frac{E[||\vfg{\tilde{\eta}}_{ij}||^2]}{\delta N}\leq \frac{C}{\delta N},
\end{align*}
according to \refmath{LindeExp}. This completes the proof of the Lindeberg condition \refmath{LindebergCond} and, thus, the proof of \ref{wts1}.\\

\vspace{1cm}
\noindent
\textit{Proof of \ref{wts2}:}\\[5mm]

At first, in order to keep the notation simple (otherwise, we would need double-indices for the diagonal elements of the hat matrix, see below) we merge the group and subject indices into one single index $\ell$ and show that  
\begin{equation}
\label{WTSConsistency}
\frac{1}{N}\sum_{\ell=1}^{N}\vf{x}_{\ell}\vf{x}_{\ell}^{\prime}\otimes (\mfg{\hat{\Sigma_{\ell}}}-\mfg{\Sigma_{\ell}}) \pconv 0
\end{equation}
holds element-wise, where $\mfg{\Sigma}_{\ell}$ and $\mfg{\hat{\Sigma}}_{\ell}$ denote the covariance matrix and its estimator for subject $\ell$, respectively, $1 \leq \ell \leq N$. Note that the MANCOVA setting (\textit{i.e.,} identical covariance matrices $\mfg{\Sigma}_{\ell}$ for all subjects within a particular group) is just a special case thereof. Due to the uniform boundedness of $\vf{x}_{\ell}$, it suffices to show that  
\begin{equation}
\label{WhiteConsistency}
\frac{1}{N}\sum_{\ell=1}^{N}(\mfg{\hat{\Sigma_{\ell}}}-\mfg{\Sigma_{\ell}}) \pconv 0
\end{equation}
holds element-wise. 
First, recall that $\vf{u}_{\ell} = \vf{Y}_{\ell} - (\vf{x}_{\ell}^{\prime}\otimes \ip)\best$. Consequently, we have
\begin{align*}
\vf{u}_{\ell}\vf{u}_{\ell}^{\prime} &= (\vf{Y}_{\ell} - (\vf{x}_{\ell}^{\prime}\otimes \ip)\best)(\vf{Y}_{\ell} - (\vf{x}_{\ell}^{\prime}\otimes \ip)\best)^{\prime} \\
&= (\vf{Y}_{\ell} - (\vf{x}_{\ell}^{\prime}(\mf{X}'\mf{X})^{-1}\mf{X}'\otimes \ip)\vf{Y})(\vf{Y}_{\ell} - (\vf{x}_{\ell}^{\prime}(\mf{X}'\mf{X})^{-1}\mf{X}'\otimes \ip)\vf{Y})^{\prime} \\
&= (\vfg{\epsilon}_{\ell} - (\vf{x}_{\ell}^{\prime}(\mf{X}'\mf{X})^{-1}\mf{X}'\otimes \ip)\vfg{\epsilon})(\vfg{\epsilon}_{\ell} - (\vf{x}_{\ell}^{\prime}(\mf{X}'\mf{X})^{-1}\mf{X}'\otimes \ip)\vfg{\epsilon})^{\prime}\\
&= \vfg{\epsilon}_{\ell} \vfg{\epsilon}_{\ell}^{\prime} - \vfg{\epsilon}_{\ell} \vfg{\epsilon}^{\prime}(\mf{X}(\mf{X}'\mf{X})^{-1}\vf{x}_{\ell}\otimes \ip) - (\vf{x}_{\ell}^{\prime}(\mf{X}'\mf{X})^{-1}\mf{X}'\otimes \ip)\vfg{\epsilon}\vfg{\epsilon}_{\ell}^{\prime} \\
&+ (\vf{x}_{\ell}^{\prime}(\mf{X}'\mf{X})^{-1}\mf{X}'\otimes \ip)\vfg{\epsilon}\vfg{\epsilon}^{\prime}(\mf{X}(\mf{X}'\mf{X})^{-1}\vf{x}_{\ell}\otimes \ip).
\end{align*} 

Therefore, we get (recalling that $p_{\ell m} = \vf{x}_{\ell}^{\prime}(\xm^{\prime}\xm)^{-1}\vf{x}_m$)
\begin{align*}
E[\vf{u}_{\ell}\vf{u}_{\ell}^{\prime}] &= E[\vfg{\epsilon}_{\ell} \vfg{\epsilon}_{\ell}^{\prime}] - \sum_{m=1}^{N}E[\vfg{\epsilon}_{\ell}\vfg{\epsilon}_m^{\prime}](p_{\ell m}\otimes \ip) - \sum_{m=1}^{N}(p_{\ell m}\otimes\ip) E[\vfg{\epsilon}_m\vfg{\epsilon}_{\ell}^{\prime}]\\
&+ \sum_{m=1}^{N}\sum_{q=1}^{N}(p_{\ell m}\otimes\ip)E[\vfg{\epsilon}_m\vfg{\epsilon}_q^{\prime}](p_{\ell q}\otimes\ip),
\end{align*}
where we have used that we can think of $X$ as a column vector containing $\vf{x}_i^{\prime}\otimes \ip$ in the $i$-th row, and $\vfg{\epsilon} = (\vfg{\epsilon}_1^{\prime},\dots,\vfg{\epsilon}_N^{\prime})^{\prime}$. Now, since $\vfg{\epsilon}_i$ and $\vfg{\epsilon}_j$ are independent for $i\neq j$, and $E[\vfg{\epsilon}_i] = \vf{0}$, the expressions from above can be further simplified to
\begin{align}
E[\vf{u}_{\ell}\vf{u}_{\ell}^{\prime}] &= E[\vfg{\epsilon}_{\ell} \vfg{\epsilon}_{\ell}^{\prime}] - E[\vfg{\epsilon}_{\ell}\vfg{\epsilon}_{\ell}^{\prime}](p_{\ell\ell}\otimes \ip) - (p_{\ell\ell}\otimes\ip) E[\vfg{\epsilon}_{\ell}\vfg{\epsilon}_{\ell}^{\prime}]+ \sum_{m=1}^{N}(p_{\ell m}\otimes\ip)E[\vfg{\epsilon}_m\vfg{\epsilon}_m^{\prime}](p_{\ell m}\otimes\ip)\nonumber\\
&= E[\vfg{\epsilon}_{\ell} \vfg{\epsilon}_{\ell}^{\prime}] - E[\vfg{\epsilon}_{\ell}\vfg{\epsilon}_{\ell}^{\prime}](p_{\ell\ell}\otimes \ip) - (p_{\ell\ell}\otimes\ip) E[\vfg{\epsilon}_{\ell}\vfg{\epsilon}_{\ell}^{\prime}]+ \sum_{m=1}^{N}(p_{\ell m}^2\otimes E[\vfg{\epsilon}_m\vfg{\epsilon}_m^{\prime}])\nonumber\\
&= E[\vfg{\epsilon}_{\ell} \vfg{\epsilon}_{\ell}^{\prime}] - p_{\ell\ell}E[\vfg{\epsilon}_{\ell}\vfg{\epsilon}_{\ell}^{\prime}] - p_{\ell\ell} E[\vfg{\epsilon}_{\ell}\vfg{\epsilon}_{\ell}^{\prime}]+ \sum_{m=1}^{N}p_{\ell m}^2 E[\vfg{\epsilon}_m\vfg{\epsilon}_m^{\prime}]\nonumber\\
&= E[\vfg{\epsilon}_{\ell} \vfg{\epsilon}_{\ell}^{\prime}] - p_{\ell\ell}E[\vfg{\epsilon}_{\ell}\vfg{\epsilon}_{\ell}^{\prime}] + \sum_{m=1}^{N}p_{\ell m}^2 (E[\vfg{\epsilon}_m\vfg{\epsilon}_m^{\prime}]-E[\vfg{\epsilon}_{\ell}\vfg{\epsilon}_{\ell}^{\prime}]),\label{errorRepresentation}
\end{align}
where the latter step was due to $p_{\ell\ell} = \sum_{m=1}^{N}p_{\ell m}^2$. Consequently, the $k$-th diagonal element of the difference $E[\vf{u}_{\ell}\vf{u}_{\ell}^{\prime}] -  \mfg{\Sigma}_{\ell}$ is
\begin{align}
\label{convergenceE}
(E[\vf{u}_{\ell}\vf{u}_{\ell}^{\prime}] -  \mfg{\Sigma}_{\ell})_{k,k}&=
\sigma_{\ell k}^{2} - p_{\ell\ell}\sigma_{\ell k}^{2} + \sum_{m=1}^{N}p_{\ell m}^2(\sigma_{mk}^{2}-\sigma_{\ell k}^{2}) - \sigma_{\ell k}^{2}  = O(N^{-1}),
\end{align}  
uniformly in $\ell$, due to the uniform boundedness assumption on the variances, and because the uniform boundedness of the covariates and the elements of $(N^{-1}\xm'\xm)^{-1}$ implies $p_{\ell m} = O(N^{-1})$. \\
Next, we consider $Var[(\vf{u}_{\ell}\vf{u}_{\ell}^{\prime} - \mfg{\Sigma}_{\ell})_{k,k}] = Var[(\vf{u}_{\ell}\vf{u}_{\ell}^{\prime})_{k,k}] = Var[u_{\ell k}^2]$. Since $\vf{u}_{\ell} = \vfg{\epsilon}_{\ell} - \sum_{m=1}^{N}p_{\ell m}\vfg{\epsilon}_m$, we have
\begin{align}
\label{VarResid}
Var[u_{\ell k}^2] = Var[\epsilon_{\ell k}^2 - 2\sum_{m=1}^{N}p_{\ell m}\epsilon_{\ell k}\epsilon_{mk} + \sum_{r,s}p_{\ell r}p_{\ell s}\epsilon_{rk}\epsilon_{sk}].
\end{align}
Using Bienayme's equality yields that \refmath{VarResid} can be further simplified to 
\begin{align}
Var[u_{\ell k}^2] &= Var[\epsilon_{\ell k}^2] \label{VarResid1}\\
&+ 4 Var[\sum_{m=1}^{N}p_{\ell m}\epsilon_{\ell k}\epsilon_{mk}]\label{VarResid2}\\
&+Var[\sum_{r,s}p_{\ell r}p_{\ell s}\epsilon_{rk}\epsilon_{sk}]\label{VarResid3}\\
&+ 2\{ Cov[\epsilon_{\ell k}^2,-2\sum_{m=1}^{N}p_{\ell m}\epsilon_{\ell k}\epsilon_{mk}]\label{VarResid4}\\
&+ Cov[\epsilon_{\ell k}^2,\sum_{r,s}p_{\ell r}p_{\ell s}\epsilon_{rk}\epsilon_{sk}]\label{VarResid5}\\
&+ Cov[-2\sum_{m=1}^{N}p_{\ell m}\epsilon_{\ell k}\epsilon_{mk},\sum_{r,s}p_{\ell r}p_{\ell s}\epsilon_{rk}\epsilon_{sk}]\}\label{VarResid6}
\end{align}
By using some algebra and the uniform boundedness conditions on $E[\epsilon_{\ell k}^{4}]$ and the covariates, one can show that \refmath{VarResid1} - \refmath{VarResid6} are at least uniformly bounded for all $\ell$ and $k$ (if not converging to $0$). Analogously, we can show that for any $\ell_1\neq \ell_2$, we have
\[
Cov[u_{\ell_1 k}^{2},u_{\ell_2 k}^{2}] = O(N^{-1}),
\]
uniformly in $\ell_1, \ell_2$ and $k$. So, summing up, we get that
\begin{align}
\label{convergenceVar}
Var\left[\frac{1}{N}\sum_{\ell = 1}^{N}u_{\ell k}^{2}\right] = \frac{1}{N^2}\left\lbrace\sum_{\ell = 1}^{N}Var[u_{\ell k}^2] + \sum_{\ell_1 \neq \ell_2}Cov[u_{\ell_1 k}^{2},u_{\ell_2 k}^{2}]\right\rbrace \overset{N \to \infty}{\longrightarrow} 0.
\end{align}
Combining \refmath{convergenceE} and \refmath{convergenceVar} and using Chebyshev's inequality, we get that
\begin{align}
\label{ConvSigmaHat}
\frac{1}{N}\sum_{\ell = 1}^{N}(u_{\ell k}^2 - \sigma_{\ell k}^{2}) \pconv 0,
\end{align}
for all $k \in \{1,2,\dots,p\}$. 

Finally, we prove that an analogous result holds for the off-diagonal elements. Define $\sigma_{\ell (jk)}:= Cov(\epsilon_{\ell j},\epsilon_{\ell k})$ for $\ell \in \{1,2,\ldots,N\}$, $j, k \in \{1,\ldots, p\}$, with $j\neq k$. Firstly, from \refmath{errorRepresentation}, one finds that
\begin{align*}
(E[\vf{u}_{\ell}\vf{u}_{\ell}^{\prime}] -  \mfg{\Sigma}_{\ell})_{j,k} &=
\sigma_{\ell (jk)} - p_{\ell\ell}\sigma_{\ell (jk)} + \sum_{m=1}^{N}p_{\ell m}^2(\sigma_{m(jk)}-\sigma_{\ell (jk)}) - \sigma_{\ell (jk)} \\
&\leq |p_{\ell\ell}\sigma_{\ell (jk)}| +\sum_{m=1}^{N}p_{\ell m}^2|\sigma_{m(jk)}-\sigma_{\ell (jk)}| \\
&= O(N^{-1}),
\end{align*}
uniformly in $\ell,j,k$, using the same arguments as before and the Cauchy-Schwarz inequality in the last step. Secondly, applying the Cauchy-Schwarz inequality again yields
\begin{align*}
Var[(\vf{u}_{\ell}\vf{u}_{\ell}^{\prime}-\mfg{\Sigma}_{\ell})_{j,k}]&= Var[(\vf{u}_{\ell}\vf{u}_{\ell}^{\prime})_{j,k}] = Var[u_{\ell j}u_{\ell k}] \leq E[u_{\ell j}^{2}u_{\ell k}^2]\\
&\leq \sqrt{E[u_{\ell j}^4]}\sqrt{E[u_{\ell k}^4]}\\
&= \sqrt{Var[u_{\ell j}^2] + E^2[u_{\ell j}^{2}]} \sqrt{Var[u_{\ell k}^2] + E^2[u_{\ell k}^2]}.
\end{align*}
According to the results from above, the latter expression is bounded from above, uniformly in $\ell$, for all $j,k$. 

\iffalse
Moreover,
\begin{align*}
Cov[u_{\ell}^{(j)}u_{\ell}^{(k)},u_m^{(j)}u_m^{(k)}] &= E[u_{\ell}^{(j)}u_{\ell}^{(k)}u_m^{(j)}u_m^{(k)}] - E[u_{\ell}^{(j)}u_{\ell}^{(k)}]E[u_m^{(j)}u_m^{(k)}]\\
&\leq |E[u_{\ell}^{(j)}u_{\ell}^{(k)}u_m^{(j)}u_m^{(k)}]| + O(N^{-1})\\
&\leq \sqrt{E[u_{\ell}^{(j)2}u_{\ell k}^2]}\sqrt{E[u_m^{(j)2}u_m^{(k)2}]} + O(N^{-1})
\end{align*}
(das $O(N^{-1})$ kommt daher, dass ja der zweite Teil, der subtrahiert wird, genau das Produkt der oben betrachteten Erwartungswerte des $(j,k)$-ten Elements ist!) Problem, das noch besteht: die oben verwendete Abschätzung für die Varianz ist wohl zu schwach!! $\rightarrow$ direkt ausrechnen (siehe Notizblaetter 1 und 2) $\rightarrow$ $Cov[u_{\ell}^{(j)}u_{\ell}^{(k)},u_m^{(j)}u_m^{(k)}]$ ist mindestens von der Ordnung $o(N^{-1})$, uniformly in $\ell,m$, passt also!\\
\fi

Moreover, one can show by tedious calculations that 
\[
Cov[u_{\ell_1 j}u_{\ell_1 k},u_{\ell_2 j}u_{\ell_2 k}] = O(N^{-1}),
\]
uniformly in $\ell_1,\ell_2$. 
Therefore, summing up, we get
\[
Var\left[\frac{1}{N}\sum_{\ell = 1}^{N}u_{\ell j}u_{\ell k}\right] = \frac{1}{N^2}\left\lbrace\sum_{\ell = 1}^{N}Var[u_{\ell j}u_{\ell k}] + \sum_{\ell_1 \neq \ell_2}Cov[u_{\ell_1 j}u_{\ell_1 k},u_{\ell_2 j}u_{\ell_2 k}]\right\rbrace \overset{N \to \infty}{\longrightarrow} 0.
\]

So, all in all, using Chebyshev's inequality, we now have convergence in probability for the off-diagonal elements, too, which completes the proof of \refmath{WhiteConsistency}. As already mentioned, \refmath{WTSConsistency} immediately follows, then, since the arguments used in the derivations above are still valid if the respective quantities are multiplied with the corresponding elements of the design matrix, due the uniform boundedness of the latter. This in turn implies that $\SigmaWhiteFull$ is weakly consistent for $\SigmaFull$, because the deterministic matrix $(N^{-1}\xm'\xm)^{-1}$ converges element-wise and has dimensions independent of $N$. This completes the proof of \ref{wts2}. \\

Finally, the statements \ref{wts1} and \ref{wts2} are combined as follows: According to the assumptions from Section 2 of the manuscript, we have $\SigmaFull \rightarrow (\mfg{\Xi}^{-1}\otimes\ip)\mfg{\Psi}(\mfg{\Xi}^{-1}\otimes\ip)$. Thus, \ref{wts2} implies that $\SigmaWhiteFull \rightarrow (\mfg{\Xi}^{-1}\otimes\ip)\mfg{\Psi}(\mfg{\Xi}^{-1}\otimes\ip)$. Since $\SigmaWhiteFull$ is of full rank for sufficiently large $N$, there is no rank jump in the convergence $\mf{\tilde{T}}\SigmaWhiteFull\mf{\tilde{T}} \pconv \mf{\tilde{T}}(\mfg{\Xi}^{-1}\otimes\ip)\mfg{\Psi}(\mfg{\Xi}^{-1}\otimes\ip)\mf{\tilde{T}}$, which implies that the convergence result also holds for the Moore-Penrose inverse. Hence, applying Slutzky's theorem to \ref{wts1} yields the final result, namely that the quadratic form
\[
Q_N(\mf{\tilde{T}}):= N (\best - \bpar)'\mf{\tilde{T}}^{\prime}(\mf{\tilde{T}}\SigmaWhiteFull \mf{\tilde{T}})^{+} \mf{\tilde{T}}(\best - \bpar)
\]
has, asymptotically, a central $\chi_f^2$ distribution, where $f = rank(\mf{\tilde{T}})$. Eventually, the statement of Theorem 1 follows by setting $\mf{\tilde{T}} = diag(\mf{T},\mf{0})$ and applying $H_0: \mf{\tilde{T}}\bpar = \mf{T}\vfg{\mu} = \vf{0}$. 

\qed

\vspace{1cm}

\subsection{Proof of Theorem 2}
\label{proof2}
{\renewcommand{\theenumi}{(B\arabic{enumi})}
\renewcommand{\labelenumi}{\theenumi}
Again, the proof comprises two main steps:
\begin{enumerate}
\item $\sqrt{N}(\vfg{\hat{\beta}}-\vfg{\beta}) \dconv \mathcal{N}(\vf{0},\mfg{\Lambda})$, where $\mfg{\Lambda}:=(\mfg{\Xi}^{-1}\otimes \ip)\mfg{\Psi}(\mfg{\Xi}^{-1}\otimes\ip)$, $\mfg{\Psi}:= \lim_{N\to \infty}(N^{-1}(\xm'\otimes\ip)\mf{S}(\xm\otimes \ip))$, $\mfg{\Xi}^{-1}:= \lim_{N \to \infty}(N^{-1}\xm'\xm)^{-1}$.\label{mancats1}
\item $(\mf{T}N\mf{\hat{D}T})^{+}\pconv (\mf{TDT})^{+}.$ \label{mancats2}
\end{enumerate}
}
\noindent
Statement \ref{mancats1} has been proven in the previous section; the proof of \ref{mancats2} works analogously to the derivations in the second part of that section, because
\[
\frac{1}{n_i}\sum_{j=1}^{n_i}\vf{u}_{ij}\vf{u}_{ij}^{\prime} - \mfg{\Sigma}_i = \frac{1}{n_i}\sum_{j=1}^{n_i}(\vf{u}_{ij}\vf{u}_{ij}^{\prime} - \mfg{\Sigma}_i) = \frac{1}{n_i}\sum_{j=1}^{n_i}(\mfg{\hat{\Sigma}}_{ij} - \mfg{\Sigma}_i).
\]
Therefore, this proof is omitted. Finally, the results are combined as follows: Let $\mf{\tilde{T}} = diag(\mf{T},\mf{0})$ and $\mf{\tilde{D}} = diag(\mf{D},\mf{I}_{cp})$. Note that actually, the particular choice of the bottom-right block of the $2\times 2$ block matrix $\mf{\tilde{D}}$ does not matter, because this part is merely inserted for a technical reason, namely in order to allow for the following calculations:  
\begin{align*}
&N(\mf{T}(\vfg{\hat{\mu}}-\vfg{\mu}))^{\prime} (\mf{TDT})^{+} \mf{T}(\vfg{\hat{\mu}}-\vfg{\mu})\\
&= N (\mf{\tilde{T}}((\vfg{\hat{\mu}}-\vfg{\mu})^{\prime},(\vfg{\hat{\nu}}-\vfg{\nu})^{\prime})^{\prime})^{\prime}diag((\mf{TDT})^{+},\mf{0})\mf{\tilde{T}}((\vfg{\hat{\mu}}-\vfg{\mu})^{\prime},(\vfg{\hat{\nu}}-\vfg{\nu})^{\prime})^{\prime}\\
&= N(\mf{\tilde{T}}(\vfg{\hat{\beta}}-\vfg{\beta}))^{\prime}(\mf{\tilde{T}\tilde{D}\tilde{T}})^{+}\mf{\tilde{T}}(\vfg{\hat{\beta}}-\vfg{\beta}).
\end{align*}

Now, according to the representation theorem and the continuous mapping theorem, statement \ref{mancats1} implies that the quadratic form
\[
N(\mf{T}(\vfg{\hat{\mu}}-\vfg{\mu}))^{\prime} (\mf{TDT})^{+} \mf{T}(\vfg{\hat{\mu}}-\vfg{\mu}) = N(\mf{\tilde{T}}(\vfg{\hat{\beta}}-\vfg{\beta}))^{\prime}(\mf{\tilde{T}\tilde{D}\tilde{T}})^{+}\mf{\tilde{T}}(\vfg{\hat{\beta}}-\vfg{\beta})
\]
has, asymptotically, the same distribution as a weighted sum $\sum_{i=1}^{a+c}\sum_{k=1}^{p}\lambda_{ik}U_{ik}$, where $U_{ik}\overset{i.i.d.}{\sim}\chi_1^2$, and the weights are the eigenvalues of the matrix of the quadratic form times the asymptotic covariance matrix of $\best$, that is, $\mf{\tilde{T}}(\mf{\tilde{T}\tilde{D}\tilde{T}})^{+}\mf{\tilde{T}}\cdot \mfg{\Lambda}$. However, since $\mf{\tilde{T}} = diag(\mf{T},\mf{0})$ and $(\mf{\tilde{T}\tilde{D}\tilde{T}})^{+} = diag((\mf{TDT})^{+},\mf{0})$, that product is equal to $diag(\mf{T}(\mf{TDT})^{+}\mf{T}\mfg{\Lambda}_{11}, \mf{0})$, where $\mfg{\Lambda}_{11}$ denotes the upper-left part of the matrix $\mfg{\Lambda}$. Thus, the weighted some from above eventually reduces to
\[
U:= \sum_{i=1}^{a}\sum_{k=1}^{p}\lambda_{ik}U_{ik},
\]
where the $\lambda_{ik}$ are the eigenvalues of $\mf{T}(\mf{TDT})^{+}\mf{T}\mfg{\Lambda}_{11}$.  Note that per definition, $\mfg{\Lambda}_{11}$ is just the asymptotic covariance matrix of the vector of adjusted means. Now, using \ref{mancats2} and Slutzky's theorem yields the desired result, namely that under $H_0:\mf{T}\vfg{\mu} = \vf{0}$, 
\[
(\mf{T}\vfg{\hat{\mu}})^{\prime} (\mf{T}\mf{\hat{D}T})^{+} \mf{T}\vfg{\hat{\mu}} = N(\mf{T}\vfg{\hat{\mu}})^{\prime} (\mf{T}N\mf{\hat{D}T})^{+} \mf{T}\vfg{\hat{\mu}} \dconv U.
\] \qed

\vspace{1cm}
\subsection{Proof of Theorem 3}
We have to show that given the data, the wild bootstrap version of the MANCATS weakly converges to the null distribution of the original MANCATS in probability, \textit{i.e.,} that given the data, 
{\renewcommand{\theenumi}{(C\arabic{enumi})}
\renewcommand{\labelenumi}{\theenumi}
\begin{enumerate}
\item $\sqrt{N}\vfg{\hat{\beta}}^{\ast} \dconv \mathcal{N}(\vf{0},\mfg{\Lambda})$ in probability, for any general linear model with coefficient vector $\vfg{\beta}$,where $\mfg{\Lambda}:=(\mfg{\Xi}^{-1}\otimes \ip)\mfg{\Psi}(\mfg{\Xi}^{-1}\otimes\ip)$.  \label{mancatswb1}
\item $(\mf{T}N\mf{\hat{D}^{\ast}T})^{+}\pconv (\mf{TDT})^{+}$ in probability. \label{mancatswb2}
\end{enumerate}
}
\noindent

\textit{Proof of \ref{mancatswb1}}:\\
\noindent

Firstly,
\begin{align*}
\sqrt{N}\hat{\vfg{\beta}}^{\ast} &= \sqrt{N}((\mf{X}'\mf{X})^{-1}\xm'\otimes\ip)\vf{Y}^{\ast} = \sqrt{N}((\mf{X}'\mf{X})^{-1}\otimes\ip)\sum_{i=1}^{a}\sum_{j=1}^{n_i}(\vf{x}_{ij}\otimes\ip) \vf{u}_{ij}T_{ij}. 
\end{align*}

So, let $\vfg{\zeta}_{ij}:=\sqrt{N}((\mf{X}'\mf{X})^{-1}\otimes\ip)(\vf{x}_{ij}\otimes\ip) \vf{u}_{ij}T_{ij}.$ Again, we use the Lindeberg-Feller theorem: $E(\vfg{\zeta}_{ij}|\vf{Y}) = 0$ is obviously fulfilled, and conditionally on the data, the $\vfg{\zeta}_{ij}$'s are independent, due to construction of the $T_{ij}$'s. Next, 
\begin{align*}
\sum_{i,j}Cov(\vfg{\zeta}_{ij}|\vf{Y}) &= \sum_{i,j}N((\mf{X}'\mf{X})^{-1}\vf{x}_{ij}\otimes \ip)Cov(\vf{u}_{ij}T_{ij}|\vf{Y}) (\vf{x}_{ij}^{\prime}(\mf{X}'\mf{X})^{-1}\otimes \ip)\\
&= \sum_{i,j}N((\mf{X}'\mf{X})^{-1}\vf{x}_{ij}\otimes \ip)\vf{u}_{ij}\vf{u}_{ij}^{\prime}Var(T_{ij}) (\vf{x}_{ij}^{\prime}(\mf{X}'\mf{X})^{-1}\otimes \ip)\\
&= N((\mf{X}'\mf{X})^{-1}\otimes\ip)\sum_{i,j}(\vf{x}_{ij}\otimes \ip)\vf{u}_{ij}\vf{u}_{ij}^{\prime} (\vf{x}_{ij}^{\prime}\otimes\ip)((\mf{X}'\mf{X})^{-1}\otimes \ip)\\
&= ((\mf{X}'\mf{X}/N)^{-1}\otimes\ip)\frac{1}{N}\sum_{i,j}(\vf{x}_{ij}\otimes \ip)\vf{u}_{ij}\vf{u}_{ij}^{\prime} (\vf{x}_{ij}^{\prime}\otimes\ip)((\mf{X}'\mf{X}/N)^{-1}\otimes \ip)\\
&\pconv (\mfg{\Xi}^{-1}\otimes \ip)\mfg{\Psi}(\mfg{\Xi}^{-1}\otimes\ip),
\end{align*}

since it has been shown in the proof of Theorem 1 that 
\[
\frac{1}{N}\sum_{i,j}(\vf{x}_{ij}\otimes \ip)\vf{u}_{ij}\vf{u}_{ij}^{\prime} (\vf{x}_{ij}^{\prime}\otimes\ip) - \frac{1}{N}\sum_{i,j}(\vf{x}_{ij}\otimes \ip)\mfg{\Sigma}_{i}(\vf{x}_{ij}^{\prime}\otimes\ip) \pconv 0.
\]

Thirdly, we check the Lindeberg condition: Doing some algebra yields
\begin{align*}
||\vfg{\zeta}_{ij}||^2 = \vfg{\zeta}_{ij}^{\prime}\vfg{\zeta}_{ij}&=\frac{1}{N}\vf{u}_{ij}^{\prime}T_{ij}(\vf{x}_{ij}^{\prime}(\mf{X'X}/N)^{-1}(\mf{X'X}/N)^{-1}\vf{x}_{ij}\otimes\ip)\vf{u}_{ij}T_{ij},
\end{align*}
Now, analogously to the proof of Theorem 1, we let $\mf{A}_{ij}:=(\vf{x}_{ij}^{\prime}(\mf{X'X}/N)^{-1}(\mf{X'X}/N)^{-1}\vf{x}_{ij}\otimes\ip)$, and we denote the $(k,l)$-th element of $A_{ij}$ by $a_{ij,k\ell}$. Consequently, we get
\begin{align}
&\sum_{i=1}^{a}\sum_{j=1}^{n_i}E[||\vfg{\zeta}_{ij}||^2 \mathbbm{1}\{||\vfg{\zeta}_{ij}||^2 > \delta\}|\vf{Y}] \nonumber\\
&= 
\frac{1}{N}\sum_{i=1}^{a}\sum_{j=1}^{n_i}E[(T_{ij}^2\vf{u}_{ij}^{\prime}\mf{A}_{ij}\vf{u}_{ij}) \mathbbm{1}\{T_{ij}^2\vf{u}_{ij}^{\prime}\mf{A}_{ij}\vf{u}_{ij} > \delta N\}|\vf{Y}] \nonumber\\
&\leq \frac{1}{N}\sum_{i,j}(\vf{u}_{ij}^{\prime}\mf{A}_{ij}\vf{u}_{ij})\sqrt{E[T_{ij}^4|\vf{Y}]}\sqrt{E[\mathbbm{1}\{T_{ij}^2\vf{u}_{ij}^{\prime}\mf{A}_{ij}\vf{u}_{ij} > \delta N\}|\vf{Y}]}.\label{WBProofLindeberg}
\end{align}
\noindent
Now, using
\[
E[\mathbbm{1}\{T_{ij}^2\vf{u}_{ij}^{\prime}\mf{A}_{ij}\vf{u}_{ij} > \delta N\}|\vf{Y}] = P(T_{ij}^4(\vf{u}_{ij}^{\prime}\mf{A}_{ij}\vf{u}_{ij})^2 > \delta^2 N^2|\vf{Y}) \leq \frac{E[T_{ij}^4|\vf{Y}](\vf{u}_{ij}^{\prime}\mf{A}_{ij}\vf{u}_{ij})^2}{\delta^2 N^2},
\]
yields 
\begin{align}
\label{WBProofLindeberg2}
&\frac{1}{N}\sum_{i,j}(\vf{u}_{ij}^{\prime}\mf{A}_{ij}\vf{u}_{ij})\sqrt{E[T_{ij}^4|\vf{Y}]}\sqrt{E[\mathbbm{1}\{T_{ij}^2\vf{u}_{ij}^{\prime}\mf{A}_{ij}\vf{u}_{ij} > \delta N\}|\vf{Y}]} \nonumber\\
&\leq \frac{1}{\delta N^2}\sum_{i,j}(\vf{u}_{ij}^{\prime}\mf{A}_{ij}\vf{u}_{ij})^2E[T_{ij}^4|\vf{Y}].
\end{align}
\noindent
Recall that the fourth moments of the wild bootstrap variables $T_{ij}$ were assumed to be uniformly bounded (see Section 3 of the manuscript).
Furthermore, using analogous arguments as in the proof of Theorem 1, one can show that $E[(\vf{u}_{ij}^{\prime}\mf{A}_{ij}\vf{u}_{ij})^2]$ is equal to the sum 
\[
\sum_{k_1=1}^{p}\sum_{\ell_1=1}^{p}\sum_{k_2=1}^{p}\sum_{\ell_2=1}^{p}a_{ij,k_1\ell_1}a_{ij,k_2\ell_2}E[u_{ijk_1}u_{ij\ell_1}u_{ijk_2}u_{ij\ell_2}].
\]
Since the fourth moments of the residuals are uniformly bounded, it follows by applying the Cauchy-Schwarz inequality repeatedly that the sum is uniformly bounded, too. Consequently, the expectation of the expression on the right handside of inequality \refmath{WBProofLindeberg2} converges to $0$. Hence, %using the monotonicity of expectations as well as inequalities \refmath{WBProofLindeberg} and \refmath{WBProofLindeberg2}, we have
\[
E\left[\sum_{i=1}^{a}\sum_{j=1}^{n_i}E[||\vfg{\zeta}_{ij}||^2 \mathbbm{1}\{||\vfg{\zeta}_{ij}||^2 > \delta\}|\vf{Y}]\right] \overset{N \to \infty}{\longrightarrow} 0.
\]
So, from the conditional version of the Markov inequality, it follows that given the data, the Lindeberg condition holds in probability. \\

\noindent

Proof of \ref{mancatswb2}:\\
\noindent

Similarly to the proof in Section \ref{proof1}, the result follows if one can show that given the data,
\begin{enumerate}[{(a)}]
\item $\lim_{n_i \to \infty}E[(n_i-c-1)^{-1}\sum_{j=1}^{n_i}(u_{ijk}^{\ast 2} - u_{ijk}^2)|\vf{Y}] \longrightarrow 0$ in probability, and
\item $\lim_{n_i \to \infty}Var[(n_i-c-1)^{-1}\sum_{j=1}^{n_i}(u_{ijk}^{\ast 2} - u_{ijk}^2)|\vf{Y}] \longrightarrow 0$ in probability,\\
\end{enumerate}
for all $i \in \{1,2,\dots,a\}$ and $k \in \{1,2,\dots,p\}$. Thereby, $u_{ijk}^{\ast}$ denotes the $k$-th component of the wild bootstrap residual vector $\vf{u}_{ij}^{\ast}:= \vf{Y}_{ij}^{\ast} - (\vf{x}_{ij}^{\prime}\otimes \ip)\best^{\ast}$ corresponding to subject $j$ in group $i$. The proof works works completely analogously as in the univariate case \citep{Zim18} and is, therefore, omitted. 

\qed  

\vspace{1cm}

\subsection{Proof of Theorem 4}
Recall that the bootstrap observations are generated according to
\[
\vf{Y_{ij}^{\star}} \overset{i.i.d.}{\sim} \mathcal{N}(\vf{0},\mfg{\hat{\Sigma}_i}), 1\leq i \leq a, 1\leq j \leq n_i,
\]
where \[
\mfg{\hat{\Sigma}_i}:=\frac{1}{n_i-c-1}\sum_{j=1}^{n_i}\vf{u_{ij}}\vf{u_{ij}^{\prime}}, 1\leq i \leq a.
\]

\iffalse
Why is this sensible? - see handwritten notes in my notebook!\\
Then, we proceed as usual: Take least squares estimates and the White covariance matrix of the bootstrap observations. Observe, however, that we do not have any covariates to consider now, due to construction of the bootstrap observations! Consequently, it turns out that the least squares estimator is just the vector of group means (no adjusted means!). Likewise, the bootstrap version of the White covariance matrix estimator is just the empirical variance, because
\[
\mfg{\hat{\Sigma}_i^{\ast}} = n_i^{-1}\sum_{j=1}^{n_i}\vf{u_{ij}}\vf{u_{ij}^{\prime}} = n_i^{-1}\sum_{j=1}^{n_i}(\vf{Y_{ij}-\bar{Y}_{i.}})(\vf{Y_{ij}-\bar{Y}_{i.}})^{\prime}.
\]
Analogously, if we take the MANCATS instead of the WTS, we just take the estimated variances in the diagonal! \\
What is interesting to see here: We do not have $n_i-1$ in the denominator; so, we could perhaps try to make some adjustment to appropriately take the df's into account when calculating the average White estimator (also, we could do this already in the PB data generation process). Check this in simulations!
\fi

Therefore, we have that given the data,
\begin{align*}
\sqrt{N}\vfg{\hat{\beta}}^{\star} \sim \mathcal{N}(\vf{0},\mfg{\hat{\Lambda}}_N),
\end{align*}
where
\[
\mfg{\hat{\Lambda}}_N = Cov(\sqrt{N}\best^{\star}|\vf{Y}) = \frac{1}{N} \sum_{i,j}((\mf{X}'\mf{X}/N)^{-1}\vf{x}_{ij}\otimes \ip)\mfg{\hat{\Sigma}}_{i} (\vf{x}_{ij}^{\prime}(\mf{X}'\mf{X}/N)^{-1}\otimes \ip).
\]

Using similar arguments as in the proof of Theorem 1 as well as in the consistency proof below, we can show that 
\[
\frac{1}{N}\sum_{i,j}(\vf{x}_{ij}\otimes \ip)\mfg{\hat{\Sigma}}_{i} (\vf{x}_{ij}^{\prime}\otimes\ip) - \frac{1}{N}\sum_{i,j}(\vf{x}_{ij}\otimes \ip)\mfg{\Sigma}_{i}(\vf{x}_{ij}^{\prime}\otimes\ip) \pconv 0,
\]
which implies
\[
\mfg{\hat{\Lambda}}_N \pconv (\mfg{\Xi}^{-1}\otimes\ip)\mfg{\Psi}(\mfg{\Xi}^{-1}\otimes\ip).
\]
Consequently, given the data in probability, the asymptotic distribution of the quadratic form 
\[
N \vfg{\hat{\mu}}^{\star\prime}\mf{T}(\mf{TDT})^{+}\mf{T}\vfg{\hat{\mu}}^{\star} = N \best^{\star\prime}\mf{\tilde{T}}(\mf{\tilde{T}}\mf{\tilde{D}}\mf{\tilde{T}})^{+}\mf{\tilde{T}}\best^{\star}
\]
is the same as the distribution of $U$ as defined in Theorem 2.   

What is left to show is that given the data, 
\begin{equation}
\label{PbConsistency}
N(\mf{\hat{D}}^{\star} - \hat{D}) \pconv \mf{0}\quad \text{in\,probability}
\end{equation}

Again, it suffices to show that given the data,
\begin{enumerate}[{(a)}]
\item $\lim_{n_i \to \infty}E[(n_i-c-1)^{-1}\sum_{j=1}^{n_i}(u_{ijk}^{\star 2} - u_{ijk}^{2})|\vf{Y}] \longrightarrow 0$ in probability, and \label{PBConsistency1}
\item $\lim_{n_i \to \infty}Var[(n_i-c-1)^{-1}\sum_{j=1}^{n_i}(u_{ijk}^{\star 2} - u_{ijk}^{2})|\vf{Y}] \longrightarrow 0$ in probability,\label{PBConsistency2}\\
\end{enumerate}
for all $i \in \{1,2,\dots,a\}$ and $k \in \{1,2,\dots,p\}$. Thereby, $u_{ijk}^{\star}$ denotes the $k$-th component of the parametric bootstrap residual vector $\vf{u}_{ij}^{\star} = (u_{ijk}^{\star})_{k=1}^{p}:= \vf{Y}_{ij}^{\star} - (\vf{x}_{ij}^{\prime}\otimes \ip)\best^{\star}$ corresponding to subject $j$ in group $i$. Firstly, observe that
\begin{equation}
u_{ijk}^{\star} = Y_{ijk}^{\star} - \sum_{i_1=1}^{a}\sum_{j_1=1}^{n_{i_1}} p_{ij,i_1j_1}Y_{i_1j_1k}^{\star},\label{BootResidRepresentation}
\end{equation}

where $p_{ij,i_1j_1}:= \vf{x}_{ij}^{\prime}(\xm'\xm)^{-1}\vf{x}_{i_1j_1}$. Consequently, we have
\begin{equation*}
u_{ijk}^{\star 2} = Y_{ijk}^{\star 2} - 2\sum_{i_1=1}^{a}\sum_{j_1=1}^{n_{i_1}} p_{ij,i_1j_1}Y_{ijk}^{\star}Y_{i_1j_1k}^{\star} + \sum_{i_2,j_2,i_3,j_3} p_{ij,i_2j_2}p_{ij,i_3j_3}Y_{i_2j_2k}^{\star}Y_{i_3j_3k}^{\star}.
\end{equation*}

Since $E[Y_{ijk}^{\star 2}|\vf{Y}] = \hat{\sigma}_{ik}^2$, and the parametric bootstrap vectors are independent, we get
\begin{align*}
0&\leq \frac{1}{n_i-c-1}\sum_{j=1}^{n_i}|E[u_{ijk}^{\star 2}|\vf{Y}]-\hat{\sigma}_{ik}^2| \\
&\leq \frac{1}{n_i-c-1}\sum_{j=1}^{n_i}2|p_{ij,ij}|\hat{\sigma}_{ik}^2 + \frac{1}{n_i-c-1}\sum_{j=1}^{n_i}\sum_{i_1=1}^{a}\sum_{j_1=1}^{n_{i_1}}p_{ij,i_1j_1}^2\hat{\sigma}_{i_1k}^2.
\end{align*}

Due to the fact that the elements of the hat matrix are bounded by $C/N$, uniformly in $i,j$, and $\hat{\sigma}_{ik}^2 \pconv \sigma_{ik}^2$, the expression above converges to $0$ in probability. So, the proof of Statement \refmath{PBConsistency1} is complete. \\
Next, we rewrite the conditional variance as follows:
\begin{align}
Var\left[\frac{1}{n_i}\sum_{j=1}^{n_i}(u_{ijk}^{\star 2}-u_{ijk}^{2})|\vf{Y}\right] &= \frac{1}{n_i^2}\sum_{j=1}^{n_i}Var[u_{ijk}^{\star 2}|\vf{Y}] + \frac{1}{n_i^2}\sum_{j\neq l}Cov[u_{ijk}^{\star 2}, u_{ilk}^{\star 2}|\vf{Y}]\nonumber 
\end{align}
Now, let
\[
A:=  \frac{1}{n_i^2}\sum_{j=1}^{n_i}Var[u_{ijk}^{\star 2}|\vf{Y}], B:=  \frac{1}{n_i^2}\sum_{j\neq \ell}Cov[u_{ijk}^{\star 2}, u_{i\ell k}^{\star 2}|\vf{Y}].
\]
If both $A$ and $B$ converge to $0$ in probability, we are done. Firstly, using \refmath{BootResidRepresentation}, Bienayme's equality and the fact that $Y_{ij}$ and $Y_{\ell m}$ are independent for $(i,j)\neq (l,m)$, we get
\begin{align*}
A &= \frac{1}{n_i^2}\sum_{j=1}^{n_i}Var[u_{ijk}^{\star 2}|\vf{Y}] \\
&= \frac{1}{n_i^2}\sum_{j=1}^{n_i}\left[Var[Y_{ijk}^{\star 2}|\vf{Y}] + 4\sum_{i_1,j_1}p_{ij,i_1j_1}^2 Var[Y_{ijk}^{\star}Y_{i_1j_1k}^{\star}|\vf{Y}]\right.\\
&+ \sum_{i_2,j_2}\sum_{i_3,j_3}p_{ij,i_2j_2}^2p_{ij,i_3j_3}^2Var[Y_{i_2j_2k}^{\star}Y_{i_3j_3k}^{\star}|\vf{Y}]\\
&+ 2\sum_{i_2,j_2}\sum_{i_3,j_3}p_{ij,i_2j_2}p_{ij,i_3j_3}Cov[Y_{i_2j_2k}^{\star}Y_{i_3j_3k}^{\star},Y_{i_2j_2k}^{\star}Y_{i_3j_3k}^{\star}|\vf{Y}] \\
&+ 2 \left( -2 p_{ij,ij} Cov[Y_{ijk}^{\star 2},Y_{ijk}^{\star 2}|\vf{Y}] + p_{ij,ij}^2  Cov[Y_{ijk}^{\star 2},Y_{ijk}^{\star 2}|\vf{Y}]\right.\\
&\left.\left.- 4\sum_{i_1,j_1}p_{ij,i_1j_1}^3 Cov[Y_{ijk}^{\star}Y_{i_1j_1k}^{\star},Y_{ijk}^{\star}Y_{i_1j_1k}^{\star}|\vf{Y}]\right)\right].
\end{align*}
Since given the data, $Y_{ijk}^{\star} \overset{ind.}{\sim}\mathcal{N}(0,\hat{\sigma}_{ik}^{2})$ for fixed $k$, we have $Var[Y_{ijk}^{\star}|\vf{Y}] = \hat{\sigma}_{ik}^{2}$ and $E[Y_{ijk}^{\star 4}|\vf{Y}] = 3\hat{\sigma}_{ik}^{4}$, for $i \in \{1,2,\ldots,a\}$. Moreover, recall that the elements of the hat matrix are uniformly bounded by $C/N$. Therefore, each term of the sum converges to $0$ in probability. For example,
\begin{align*}
\sum_{i_2,j_2}\sum_{i_3,j_3}p_{ij,i_2j_2}p_{ij,i_3j_3}Cov[Y_{i_2j_2k}^{\star}Y_{i_3j_3k}^{\star},Y_{i_2j_2k}^{\star}Y_{i_3j_3k}^{\star}|\vf{Y}]&= \sum_{i_2,j_2}\sum_{i_3,j_3}p_{ij,i_2j_2}p_{ij,i_3j_3}Var[Y_{i_2j_2k}^{\star}Y_{i_3j_3k}^{\star}|\vf{Y}]\\ 
&\leq \frac{3C^2}{N^2}\sum_{i_2,j_2}\sum_{i_3,j_3}\hat{\sigma}_{i_2k}^{2}\hat{\sigma}_{i_3k}^{2}\\
&= \frac{3C^2}{N^2}\sum_{i_2=1}^{a}n_{i_2}\hat{\sigma}_{i_2k}^{2}\sum_{i_3=1}^{a}n_{i_3}\hat{\sigma}_{i_3k}^{2}\\
&= 3C^2 \sum_{i_2=1}^{a}\frac{n_{i_2}}{N}\hat{\sigma}_{i_2k}^{2}\sum_{i_3=1}^{a}\frac{n_{i_3}}{N}\hat{\sigma}_{i_3k}^{2},
\end{align*} 
and consequently,
\begin{align*}
\frac{1}{n_i^2}\sum_{j=1}^{n_i}\sum_{i_2,j_2}\sum_{i_3,j_3}p_{ij,i_2j_2}p_{ij,i_3j_3}Var[Y_{i_2j_2k}^{\star}Y_{i_3j_3k}^{\star}|\vf{Y}] &\leq \frac{3C^2}{n_i} \sum_{i_2=1}^{a}\frac{n_{i_2}}{N}\hat{\sigma}_{i_2k}^{2}\sum_{i_3=1}^{a}\frac{n_{i_3}}{N}\hat{\sigma}_{i_3k}^{2}\pconv 0,
\end{align*}
since $\frac{n_{i_2}}{N}\hat{\sigma}_{i_2k}^{2} \pconv \kappa_{i_2}^{-1}\sigma_{i_2k}^{2}$, $\frac{n_{i_3}}{N}\hat{\sigma}_{i_3k}^{2} \pconv \kappa_{i_3}^{-1}\sigma_{i_3k}^{2}$. The proof for the other summands of $A$ works analogously and is, therefore, omitted. Likewise, by using the bilinearity of the covariance and considering each part of the resulting sum separately, it follows that $B \pconv 0$. So, summing up, the proof of Statement \refmath{PBConsistency2} and, thus, of Theorem 4 is complete. 

\qed

\clearpage

\section{Further simulation results: Type I error rates}
\label{Supp:typeI}

\begin{table}[h!t]

\caption{Empirical type I error rates (in \%) of the Wald-type test (WT), the wild bootstrap MANCATS test (MW), and the parametric bootstrap MANCATS test (MP) for $a=2$ groups, $p=8$ dimensions, $c=2$ fixed covariates, and the following homoskedastic singular covariance scenario: $\mfg{\Sigma}_i = (\vf{s}_1,\ldots,\vf{s}_8)$, where $\vf{s}_1 = \ldots = \vf{s}_5 = (1,1,1,1,1,1/2,1/2,1/2)'$, $\vf{s}_6 = (1/2,1/2,1/2,1/2,1/2,1,1/2,1/2)'$, $\vf{s}_7 = (1/2,1/2,1/2,1/2,1/2,1/2,1,1/2)'$, $\vf{s}_8 = (1/2,1/2,1/2,1/2,1/2,1/2,1/2,1)'$, $i \in \{1,2\}$. Errors were drawn either from a standard normal, a Chi-square(5), a standard lognormal, or a double exponential distribution.}
\vspace{3mm}
\begin{tabular}{*{13}{r}}
 \hspace{0.2cm} & \multicolumn{3}{c}{\textbf{Normal}} & \multicolumn{3}{c}{\textbf{Chi-square(5)}} & \multicolumn{3}{c}{\textbf{Lognormal}}& \multicolumn{3}{c}{\textbf{Double exp.}}\\
$(n_1,n_2)$ & WT& MW& MP & WT& MW& MP & WT& MW& MP & WT& MW& MP \\
\hline
$(20,20)$ & $4.2$ & $6.1$ & $5.0$ &$4.1$&$6.4$&$5.1$&$2.7$&$6.6$&$5.3$&$4.2$&$6.1$&$5.1$  \\
$(10,10)$ & $13.9$ & $7.6$ & $5.2$ & $12.6$ & $7.3$ & $5.3$ & $9.8$ & $5.9$ & $4.5$ & $13.0$ & $7.0$ & $4.9$  \\
$(10,20)$ &$13.2$ & $8.2$ & $5.4$ &$11.9$&$7.7$&$5.3$&$8.2$&$6.9$&$5.8$&$11.7$&$7.2$&$4.5$\\
$(20,10)$ &$12.5$ & $8.0$ & $5.1$ &$11.0$&$7.9$&$5.5$&$7.6$&$7.3$&$4.7$&$11.2$&$7.1$&$4.9$ \\
\label{Table:TypeI2gr8dim}
\end{tabular}  
\end{table}

\begin{table}[h!t]

\caption{Empirical type I error rates (in \%) of the Wald-type test (WT), the wild bootstrap MANCATS test (MW), and the parametric bootstrap MANCATS test (MP) for $a=2$ groups, $p=2$ dimensions, and the following homoskedastic singular covariance scenario: $\mfg{\Sigma}_i = diag(1,0.25) + 0.5(J_2-I_2)$, $i \in \{1,2\}$. The values of the first covariate were equally spaced between $-10$ and $10$, whereas the second covariate was a sample from a standard normal or lognormal distribution. Errors were drawn either from a standard normal, a Chi-square(5), a standard lognormal, or a double exponential distribution.}  
\vspace{3mm}
\begin{tabular}{cc@{\hspace{8pt}}c@{\hspace{8pt}}c@{\hspace{8pt}}c@{\hspace{8pt}}c@{\hspace{8pt}}c@{\hspace{8pt}}c@{\hspace{8pt}}c@{\hspace{8pt}}c@{\hspace{8pt}}c@{\hspace{8pt}}c@{\hspace{8pt}}c@{\hspace{8pt}}c@{\hspace{8pt}}}
\hspace{0.1cm} & \hspace{0.2cm} & \multicolumn{3}{c}{\textbf{Normal}} & \multicolumn{3}{c}{\textbf{Chi-square(5)}} & \multicolumn{3}{c}{\textbf{Lognormal}}& \multicolumn{3}{c}{\textbf{Double exp.}}\\
Covariate& $(n_1,n_2)$ & WT& MW& MP & WT& MW& MP & WT& MW& MP & WT& MW& MP \\
\hline
\multirow{4}{*}{Normal} & $(20,20)$ & $2.2$ & $5.8$ & $5.0$ &$1.8$&$5.6$&$4.8$&$1.3$&$4.8$&$4.6$&$2.0$&$5.7$&$4.8$  \\
 &$(10,10)$ & $2.9$ & $7.0$ & $4.7$ & $2.7$ & $6.8$ & $5.2$ & $1.7$ & $4.8$ & $4.3$ & $2.8$ & $6.7$ & $4.9$  \\
 &$(10,20)$ &$2.4$ & $6.2$ & $5.2$ &$2.5$&$6.1$&$5.4$&$1.3$&$4.6$&$4.6$&$2.3$&$6.1$&$5.3$\\
 &$(20,10)$ &$2.2$ & $5.6$ & $5.1$ &$2.6$&$5.9$&$5.3$&$1.5$&$4.6$&$4.7$&$2.2$&$5.5$&$4.8$ \\
\hline 
 \multirow{4}{*}{Lognormal} & $(20,20)$ & $1.9$ & $5.9$ & $5.0$ &$1.6$&$5.3$&$4.8$&$1.1$&$4.5$&$4.7$&$1.8$&$5.3$&$4.8$  \\
 &$(10,10)$ & $2.4$ & $6.4$ & $4.8$ & $2.3$ & $6.5$ & $5.1$ & $1.6$ & $4.9$ & $4.8$ & $2.5$ & $6.5$ & $5.2$  \\
 &$(10,20)$ &$2.5$ & $5.9$ & $5.7$ &$2.4$&$6.0$&$5.8$&$1.4$&$4.6$&$5.4$&$2.4$&$6.1$&$5.9$\\
 &$(20,10)$ &$2.0$ & $5.4$ & $4.8$ &$2.1$&$6.0$&$4.9$&$1.2$&$4.6$&$4.6$&$1.9$&$5.4$&$4.7$ \\
\label{Table:TypeIrandcovnormal}
\end{tabular}  
\end{table}

\begin{table}[h!t]

\caption{Empirical type I error rates (in \%) of the Wilks' Lambda test (WI), the Wald-type test (WT), the wild bootstrap MANCATS test (MW), and the parametric bootstrap MANCATS test (MP) for $a=2$ groups, $p=2$ dimensions, $c=2$ fixed covariates, and the following heteroskedastic covariance scenario: $\mfg{\Sigma}_i = i\cdot I_2 + 0.5(J_2-I_2)$, $i \in \{1,2\}$. Errors were drawn either from a standard normal, a Chi-square(5), a standard lognormal, or a double exponential distribution. Within each of the $n_{sim} = 1 000$ simulation runs, $n_{boot} = 500$ bootstrap samples were drawn.}
\vspace{3mm}
\begin{tabular}{cc@{\hspace{5pt}}c@{\hspace{5pt}}c@{\hspace{5pt}}c@{\hspace{5pt}}c@{\hspace{5pt}}c@{\hspace{5pt}}c@{\hspace{5pt}}c@{\hspace{5pt}}c@{\hspace{5pt}}c@{\hspace{5pt}}c@{\hspace{5pt}}c@{\hspace{5pt}}c@{\hspace{5pt}}c@{\hspace{5pt}}c@{\hspace{5pt}}c@{\hspace{5pt}}}
\hspace{0.2cm} & \multicolumn{4}{c}{\textbf{Normal}} & \multicolumn{4}{c}{\textbf{Chi-square(5)}} & \multicolumn{4}{c}{\textbf{Lognormal}}& \multicolumn{4}{c}{\textbf{Double exp.}}\\
$(n_1,n_2)$ & WI & WT& MW& MP & WI & WT& MW& MP & WI & WT& MW& MP & WI & WT& MW& MP \\
\hline
$(200,200)$ & $2.9$ & $5.8$ & $5.3$ & $5.4$ & $1.8$&$5.1$&$4.5$&$4.5$& $2.7$ & $4.8$&$6.0$&$6.2$& $1.8$ & $6.0$&$5.4$&$5.0$  \\
 $(100,100)$ & $2.4$ & $5.9$ & $5.2$ & $4.1$ & $3.3$& $6.6$ & $7.7$ & $6.9$ & $2.6$ & $5.0$ & $6.1$ & $4.5$ & $2.2$& $6.8$ & $5.6$ & $5.5$  \\
 $(150,50)$ & $9.4$ &$6.2$ & $4.1$ & $3.8$ & $9.6$ & $5.3$&$5.4$&$4.9$& $9.3$ & $4.9$&$5.1$&$6.2$& $9.7$ &$6.1$&$5.7$&$4.1$\\
 $(50,150)$ & $4.5$&$5.4$ & $5.4$ & $5.6$ &$4.0$&$6.9$&$7.0$&$6.5$& $3.7$ & $6.4$&$5.2$&$5.2$& $3.7$ & $6.4$&$5.7$&$5.7$ \\
\label{Table:TypeIlargeN}
\end{tabular}  
\end{table}

\pagebreak

\section{Further simulation results: Achieved power}
\label{Supp:achieved}

\begin{figure}[h!b]
\centering
\includegraphics[width = 14cm, scale = 1]{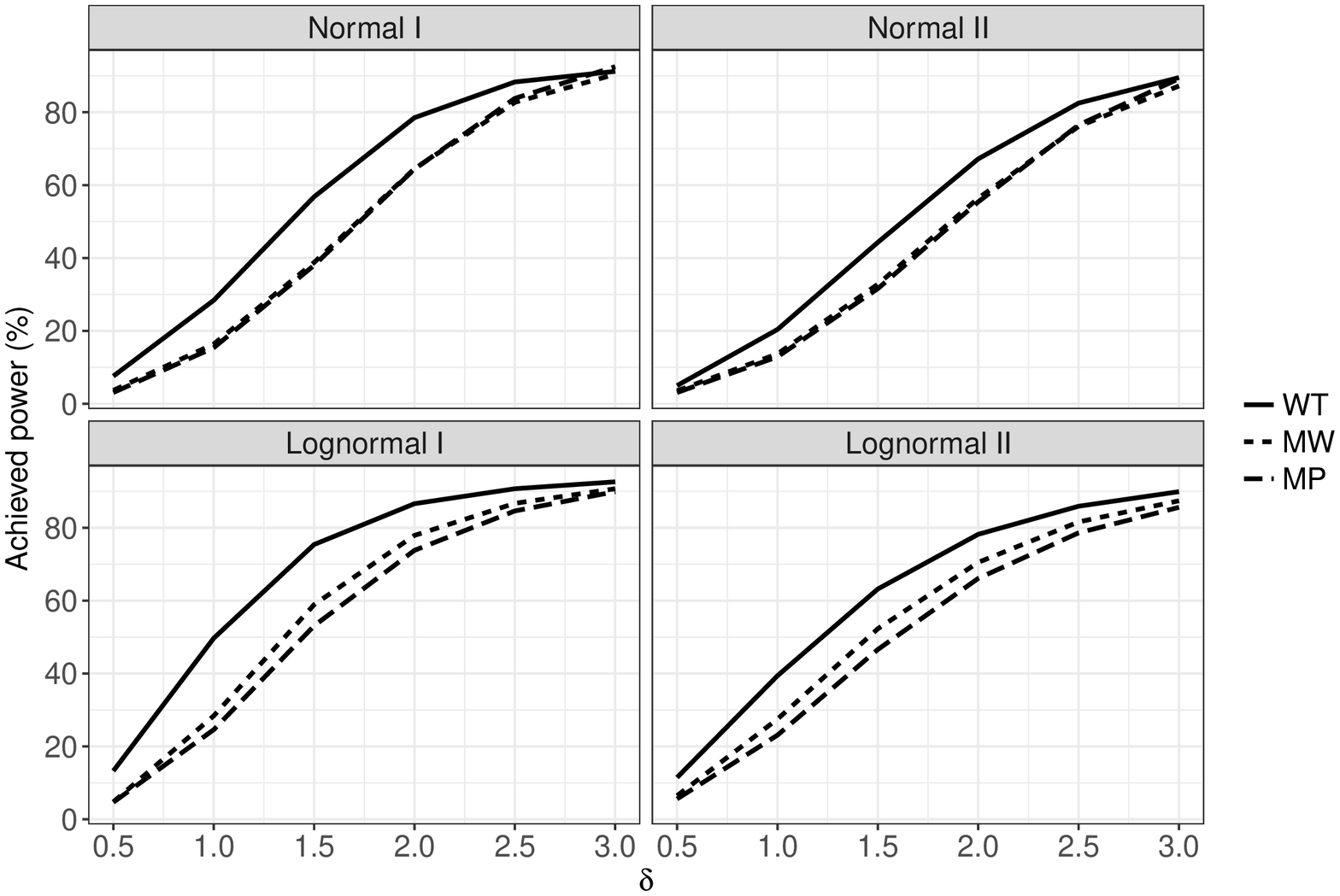}
\caption{Achieved power (\textit{i.e.}, empirical power minus type I error rate) of the Wald-type test (WT), the wild bootstrap MANCATS test (MW), and the parametric bootstrap MANCATS test (MP) for a bivariate outcome ($p=2$) and $c = 2$ covariates. Data were generated for a setting with $a=2$ groups of sizes $n_1 = n_2 = 20$, with $\vfg{\mu}_1=(0,0)'$, $\vfg{\mu}_2 = (\delta,0)'$, and choices of covariance matrices I: $\mfg{\Sigma}_i = I_2 + 0.5(J_2-I_2)$, II: $\mfg{\Sigma}_i = i\cdot I_2 + 0.5(J_2-I_2)$, $i \in \{1,2\}$. Errors were drawn from a standard normal or lognormal distribution, respectively.\label{Figure:Power1achieved}}
\end{figure}

\begin{figure}[h!b]
\centering
\includegraphics[width = 14cm, scale = 1]{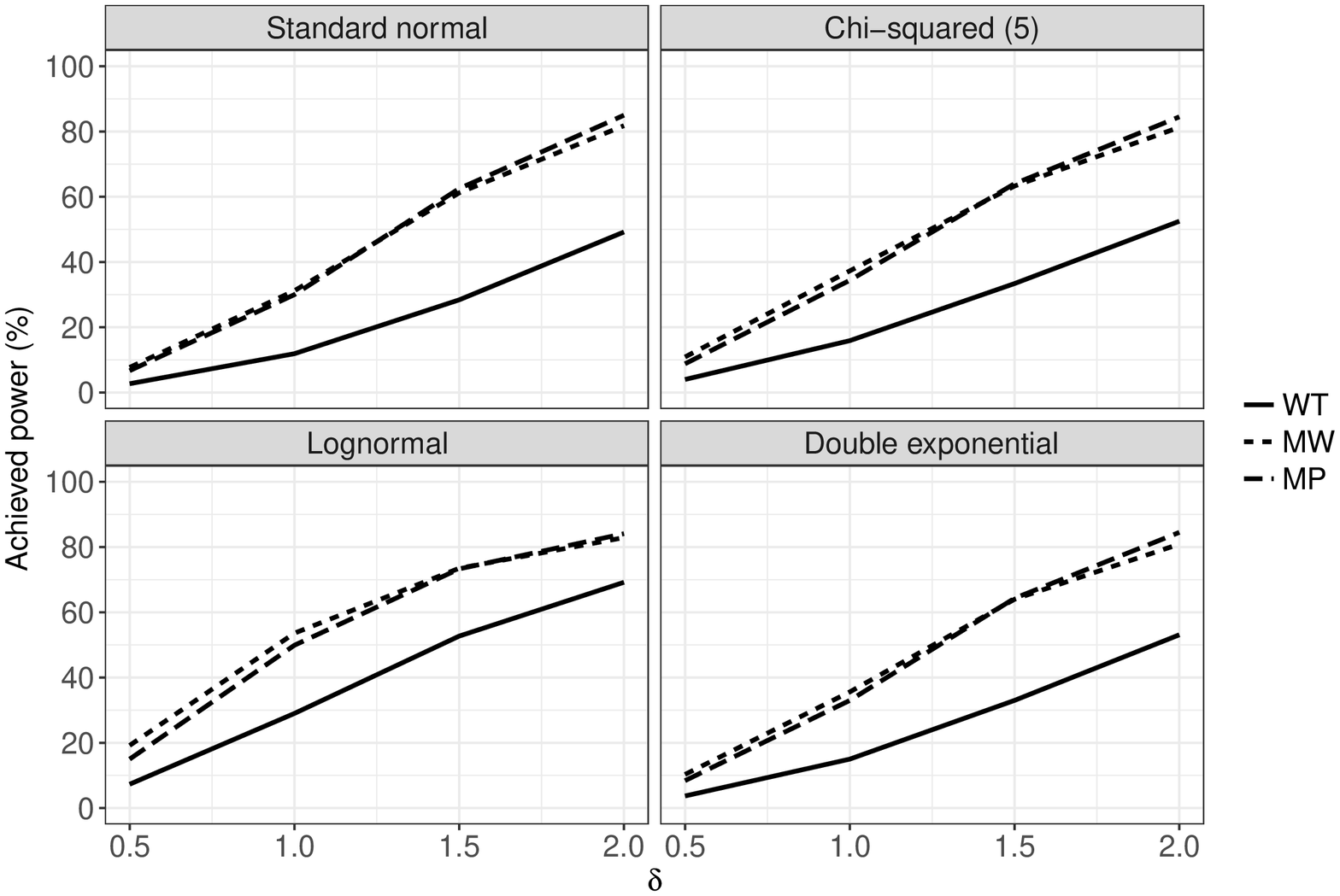}
\caption{Achieved power (\textit{i.e.}, empirical power minus type I error rate) of the Wald-type test (WT), the wild bootstrap MANCATS test (MW), and the parametric bootstrap MANCATS test (MP) for a bivariate outcome ($p=2$) and $c = 2$ covariates. Data were generated for a setting with $a=2$ groups of sizes $n_1 = n_2 = 10$, with $\vfg{\mu}_1=(0,0)'$, $\vfg{\mu}_2 = (\delta,\delta)'$, assuming the singular covariance matrix structure III: $\mfg{\Sigma}_i = diag(1,0.25) + 0.5(J_2-I_2)$, $i \in \{1,2\}$. Errors were drawn from a standard normal, a Chi-square(5), a standard lognormal, or a double exponential distribution, respectively.\label{Figure:Power2achieved}}
\end{figure}

\clearpage

\bibliography{MancovaBibl}

\begin{thebibliography}{30}
\providecommand{\natexlab}[1]{#1}
\providecommand{\url}[1]{\texttt{#1}}
\expandafter\ifx\csname urlstyle\endcsname\relax
  \providecommand{\doi}[1]{doi: #1}\else
  \providecommand{\doi}{doi: \begingroup \urlstyle{rm}\Url}\fi

\bibitem[Anderson(2003)]{And}
TW~Anderson.
\newblock \emph{An Introduction to Multivariate Statistical Analysis}.
\newblock John Wiley \& Sons, Hoboken, New Jersey, 3rd edition, 2003.

\bibitem[Arellano(1987)]{Are87}
M~Arellano.
\newblock Computing robust standard errors for within-group estimators.
\newblock \emph{Oxford Bulletin of Economics and Statistics}, 49:\penalty0
  431--434, 1987.

\bibitem[Brunner et~al.(1997)Brunner, Dette, and Munk]{Bru97}
E~Brunner, H~Dette, and A~Munk.
\newblock Box-type approximations in nonparametric factorial designs.
\newblock \emph{Journal of the American Statistical Association}, 92:\penalty0
  1494--1502, 1997.

\bibitem[Cameron et~al.(2008)Cameron, Gelbach, and Miller]{Cam08}
AC~Cameron, JB~Gelbach, and DL~Miller.
\newblock Bootstrap-based improvements for inference with clustered errors.
\newblock \emph{The Review of Economics and Statistics}, 90\penalty0
  (3):\penalty0 414--427, 2008.

\bibitem[Cribari-Neto(2004)]{Cri}
F~Cribari-Neto.
\newblock Asymptotic inference under heteroskedasticity of unknown form.
\newblock \emph{Computational Statistics and Data Analysis}, 45:\penalty0
  215--233, 2004.

\bibitem[Fan and Zhang(2017)]{Fan17}
C~Fan and D~Zhang.
\newblock Rank repeated measures analysis of covariance.
\newblock \emph{Communications in Statistics - Theory and Methods}, 46\penalty0
  (3):\penalty0 1158--1183, 2017.

\bibitem[Freire et~al.(2018)Freire, Ferradás, Nunez, and Valle]{Fre18}
C~Freire, MDM Ferradás, JC~Nunez, and A~Valle.
\newblock Coping flexibility and eudaimonic well-being in university students.
\newblock \emph{Scandinavian Journal of Psychology}, 59\penalty0 (4):\penalty0
  433--442, 2018.

\bibitem[Friedrich and Pauly(2018)]{Fri18}
S~Friedrich and M~Pauly.
\newblock {MATS:} inference for potentially singular and heteroscedastic
  {MANOVA}.
\newblock \emph{Journal of Multivariate Analysis}, 165:\penalty0 166--179,
  2018.

\bibitem[Friedrich et~al.(2017)Friedrich, Konietschke, and Pauly]{Fri17}
S~Friedrich, F~Konietschke, and M~Pauly.
\newblock A wild bootstrap approach for nonparametric repeated measurements.
\newblock \emph{Computational Statistics and Data Analysis}, 113:\penalty0
  38–--52, 2017.

\bibitem[Huitema(2011)]{Hui}
BE~Huitema.
\newblock \emph{The Analysis of Covariance and Alternatives: Statistical
  Methods for Experiments, Quasi-Experiments, and Single-Case Studies}.
\newblock Wiley, New York, 2011.

\bibitem[Hyams et~al.(2018)Hyams, Hay-McCutcheon, and Scogin]{Hya18}
AV~Hyams, M~Hay-McCutcheon, and F~Scogin.
\newblock Hearing and quality of life in older adults.
\newblock \emph{Journal of Clinical Psychology}, 74\penalty0 (10):\penalty0
  1874--1883, 2018.

\bibitem[{International Council for Harmonisation of Technical Requirements for
  Pharmaceuticals for Human Use}(1998)]{IchE9}
{International Council for Harmonisation of Technical Requirements for
  Pharmaceuticals for Human Use}.
\newblock {ICH} harmonized tripartite guideline: Statistical principles for
  clinical trials {E9}, step 4, 1998.

\bibitem[Jackson et~al.(2018)Jackson, Jones, Hsu, Stafstrom, Lin, Almane,
  Koehn, Seidenberg, and Hermann]{Jac18}
DC~Jackson, JE~Jones, DA~Hsu, CE~Stafstrom, JJ~Lin, D~Almane, MA~Koehn,
  M~Seidenberg, and BP~Hermann.
\newblock Language function in childhood idiopathic epilepsy syndromes.
\newblock \emph{Brain and Language}, Accepted, 2018.

\bibitem[Konietschke et~al.(2015)Konietschke, Bathke, Harrar, and Pauly]{Kon15}
F~Konietschke, AC~Bathke, SW~Harrar, and M~Pauly.
\newblock Parametric and nonparametric bootstrap methods for general {MANOVA}.
\newblock \emph{Journal of Multivariate Analysis}, 140:\penalty0 291--301,
  2015.

\bibitem[Liu(1988)]{Liu}
RY~Liu.
\newblock Bootstrap procedures under some non-i.i.d. models.
\newblock \emph{The Annals Of Statistics}, 16\penalty0 (4):\penalty0
  1696--1708, 1988.

\bibitem[Lyndon et~al.(2017)Lyndon, Henning, Alyami, Krishna, Yu, and
  Hill]{Lyn17}
MP~Lyndon, MA~Henning, H~Alyami, S~Krishna, TC~Yu, and AG~Hill.
\newblock The impact of a revised curriculum on academic motivation, burnout,
  and quality of life among medical students.
\newblock \emph{Journal of Medical Education and Curricular Development}, 4,
  2017.

\bibitem[MacKinnon(2012)]{Mac10}
JG~MacKinnon.
\newblock Thirty years of heteroskedasticity-robust inference.
\newblock \emph{Queen's Economic Department Working Paper}, 1268, 2012.
\newblock URL
  \url{http://qed.econ.queensu.ca/working\_papers/papers/qed\_wp\_1268.pdf}.

\bibitem[MacKinnon and White(1985)]{Mac}
JG~MacKinnon and H~White.
\newblock Some heteroskedasticity-consistent covariance matrix estimators with
  improved finite sample properties.
\newblock \emph{Journal of Econometrics}, 29:\penalty0 305--325, 1985.

\bibitem[Mammen(1993)]{Mam}
E~Mammen.
\newblock Bootstrap and wild bootstrap for high dimensional linear models.
\newblock \emph{The Annals of Statistics}, 21\penalty0 (1):\penalty0 255--285,
  1993.

\bibitem[Memarmoghaddam et~al.(2016)Memarmoghaddam, Torbati, Sohrabi, Mashhadi,
  and Kashi]{Mem16}
M~Memarmoghaddam, HT~Torbati, M~Sohrabi, A~Mashhadi, and A~Kashi.
\newblock Effects of a selected exercise programon executive function of
  children with attention deficit hyperactivity disorder.
\newblock \emph{Journal of Medicine and Life}, 9\penalty0 (4):\penalty0
  373--379, 2016.

\bibitem[{R Development Core Team}(2018)]{Rco}
{R Development Core Team}.
\newblock R: A language and environment for statistical computing, 2018.
\newblock URL \url{http://www.R-project.org}.
\newblock R Foundation for Statistical Computing, Vienna, Austria.

\bibitem[Rencher and Christensen(2002)]{Ren}
AC~Rencher and WF~Christensen.
\newblock \emph{Methods of multivariate analysis}.
\newblock John Wiley \& Sons, Hoboken, New Yersey, 3rd edition, 2002.

\bibitem[Roldan-Valadez et~al.(2013)Roldan-Valadez, Pina-Jimenez, Favila, and
  Rios]{Rol13}
E~Roldan-Valadez, C~Pina-Jimenez, R~Favila, and C~Rios.
\newblock Gender and age groups interactions in the quantification of bone
  marrow fat content in lumbar spine using {3T MR} spectroscopy: a multivariate
  analysis of covariance ({MANCOVA}).
\newblock \emph{European Journal of Radiology}, 82\penalty0 (11):\penalty0
  697--702, 2013.

\bibitem[Setyowibowo et~al.(2018)Setyowibowo, Purba, Hunfeld, Iskandarsyah,
  Sadarjoen, Passchier, and Sijbrandij]{Set18}
H~Setyowibowo, FD~Purba, JAM Hunfeld, A~Iskandarsyah, SS~Sadarjoen,
  J~Passchier, and M~Sijbrandij.
\newblock Quality of life and health status of indonesian women with breast
  cancer symptoms before the definitive diagnosis: A comparison with indonesian
  women in general.
\newblock \emph{{PLoS One}}, 13\penalty0 (7):\penalty0 e0200966, 2018.

\bibitem[Sullivan and D'Agostino(2003)]{Sul03}
LM~Sullivan and RB~D'Agostino.
\newblock Robustness and power of analysis of covariance applied to ordinal
  scaled data as arising in randomized controlled trials.
\newblock \emph{Statistics in Medicine}, 22:\penalty0 1317--1334, 2003.

\bibitem[Timm(2002)]{Tim}
NH~Timm.
\newblock \emph{Applied Multivariate Analysis}.
\newblock Springer, New York, Berlin, Heidelberg, 2002.

\bibitem[Tournikioti et~al.(2018)Tournikioti, Ferentinos, Michopoulos, Dikeos,
  Soldatos, and Douzenis]{Tou18}
K~Tournikioti, P~Ferentinos, I~Michopoulos, D~Dikeos, CR~Soldatos, and
  A~Douzenis.
\newblock Sex-related variation of neurocognitive functioning in bipolar
  disorder: Focus on visual memory and associative learning.
\newblock \emph{Psychiatry Research}, 267:\penalty0 499--505, 2018.

\bibitem[White(1980)]{Whi}
H~White.
\newblock A heteroskedasticity-consistent covariance matrix estimator and a
  direct test for heteroskedasticity.
\newblock \emph{Econometrica}, 48:\penalty0 817--838, 1980.

\bibitem[Wu(1986)]{Wu}
CFJ Wu.
\newblock Jackknife, bootstrap and other resampling methods in regression
  analysis.
\newblock \emph{The Annals of Statistics}, 14\penalty0 (4):\penalty0
  1261--1295, 1986.

\bibitem[Zimmermann et~al.(2019)Zimmermann, Pauly, and Bathke]{Zim18}
G~Zimmermann, M~Pauly, and AC~Bathke.
\newblock Small-sample performance and underlying assumptions of a
  bootstrap-based inference method for a general analysis of covariance model
  with possibly heteroskedastic and nonnormal errors.
\newblock \emph{Statistical Methods in Medical Research}, Accepted, 2019.

\end{thebibliography}

\end{document}